\documentclass[acmsmall]{acmart}

\AtBeginDocument{%
  \providecommand\BibTeX{{%
    \normalfont B\kern-0.5em{\scshape i\kern-0.25em b}\kern-0.8em\TeX}}}

\usepackage{graphicx} 
\usepackage{comment}
\usepackage{graphics}
\usepackage{multirow}
\usepackage{url}
\usepackage{subfigure}

\usepackage{algorithmic}
\usepackage{threeparttable}
\usepackage{footnote} 
\usepackage{hyperref} 
\usepackage{multirow} 
\usepackage{amsmath} 
\usepackage{amsfonts} 
\usepackage{amsthm} 

\usepackage{pgfplots}
\usetikzlibrary{matrix}

\usepackage{color}

\usepgfplotslibrary{groupplots}

\usepackage[linesnumbered,ruled,vlined]{algorithm2e}
\SetKwInput{KwInput}{Input}                
\SetKwInput{KwOutput}{Output}              

\pgfplotsset{compat=newest}

\usepackage{tikz}
\usetikzlibrary{positioning}

\usepackage{subfigure}
\usepackage{comment}

\acmJournal{TWEB}

\begin{document}

\title{Causality and Correlation Graph Modeling for Effective and Explainable Session-based Recommendation}

\author{Huizi Wu}
\authornote{The first two authors contributed equally.}
\affiliation{
  \institution{RIIS $\&$ SIME, Shanghai University of Finance and Economics}
  \city{Shanghai}
  \country{China}}
\email{wuhuizisufe@gmail.com}

\author{Cong Geng}
\authornotemark[1]
\affiliation{
  \institution{RIIS $\&$ SIME, Shanghai University of Finance and Economics}
  \city{Shanghai}
  \country{China}}
\email{gcong.leslie@gmail.com}

\author{Hui Fang}
\authornote{Corresponding author.}
\affiliation{
  \institution{RIIS $\&$ SIME, Shanghai University of Finance and Economics}
  \city{Shanghai}
  \country{China}}
\email{fang.hui@mail.shufe.edu.cn}

\begin{abstract}
Session-based recommendation which has been witnessed a booming interest recently, focuses on predicting a user's next interested item(s) based on an anonymous session. Most existing studies adopt complex deep learning techniques (e.g., graph neural networks) for effective session-based recommendation. However, they merely address \emph{co-occurrence} between items, but fail to well distinguish \emph{causality} and \emph{correlation} relationship. Considering the varied interpretations and characteristics of causality and correlation relationship between items, 
in this study, we propose a novel method denoted as CGSR by jointly modeling causality and correlation relationship between items. In particular, we construct cause, effect and correlation graphs from sessions by simultaneously considering the false causality problem. We further design a graph neural network-based method for session-based recommendation. To conclude, we strive to explore the relationship between items from specific ``causality" (directed) and ``correlation" (undirected) perspectives. Extensive experiments on three datasets show that our model outperforms other state-of-the-art methods in terms of recommendation accuracy. Moreover, we further propose an explainable framework on CGSR, and demonstrate the explainability of our model via case studies on Amazon dataset.
\end{abstract}



\maketitle

\section{Introduction}\label{sec:introduction}

Session-based recommendation (SR) has attracted wide attention in recent years~\cite{fang2020deep}. In contrast to traditional recommendation modeling users' static preferences, it processes time-aware user-item interactions to capture the dynamic preferences.
Its task is to recommend next item a user will probably like given an anonymous session.
Quite a series of models have been proposed to improve the performance of SR, ranging from the early Markov Chain-based ones \cite{rendle2010factorizing} to the recent deep learning-based ones, including recurrent neural network (RNN)-based \cite{dobrovolny2021session,leng2021hierarchical}, attention mechanism-based \cite{zhang2021personal,kang2018self} and graph neural network (GNN)-based methods \cite{zhou2021temporal,yu2021graph}.

Although some algorithms have obtained encouraging improvements as reported, they still suffer from the following limitations: (1) Most RNN-based and attention-based methods \cite{li2017neural,zhang2021personal} focus on the dependency relationship of items within a session, but fail to easily capture item transitions across sessions;
(2) The general GNN-based methods \cite{chen2020handling} alleviate the aforementioned issue by constructing session graph across sessions, but they mainly model \emph{correlation} relationship (namely co-occurrence) between items. Thus, similar to RNN-based and attention-based methods, they neglect to well distinguish directed relationship from undirected relationship between items.
\begin{figure}[htb]
    \centering
    \includegraphics[width=9cm]{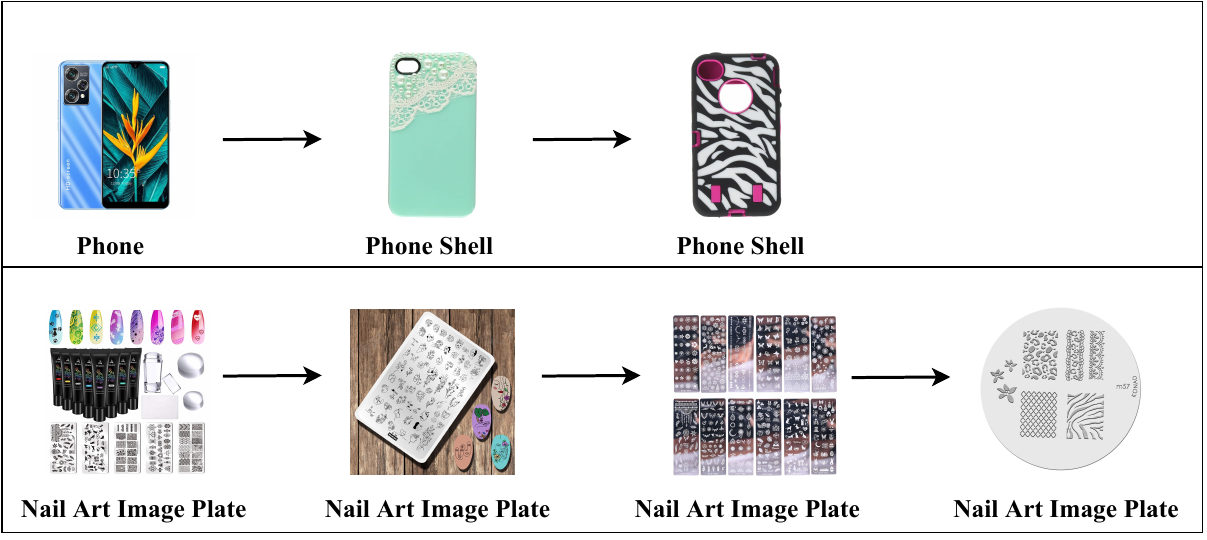}\vspace{-1mm}
    \caption{Two session cases from a real dataset.}\vspace{-1mm}
    \label{fig:example1}
\end{figure}
In this study, we call this directed relationship as ``causality", whilst the undirected relationship as ``correlation”. Figure 1 illustrates an example about two sessions from a real dataset. As we can see, there exist two sessions containing interacted items in chronological order. In the first session, the phone leads the user to buy the phone shell, whilst different nail art image plates in the second session indicate the correlation relationship between them. In this case, it is worthwhile to explore different relationship among items in terms of causality and correlation perspective for more effective session-based recommendation.

Besides, ``causality" refers to a direct and asymmetric relationship, which is calculated from a relatively large volume of sessions, and particularly considered for session-based recommendation scenario. It is not strictly equivalent to that in the traditional sense, like causal modeling by econometric approaches.
Here, the higher the edge weight in causality graph, the higher the probability of this directional relationship between items. Moreover, ``causality" relates to both cause and effect where in recommendation the cause item is partly responsible for the effect item, meanwhile the effect is partly dependent on the cause (cause$\rightarrow$effect) \cite{yao2020survey,scholkopf2019causality}. Our model divides it into two parts since we strive to maximize item transition from cause $\rightarrow$ effect whilst minimize that from effect $\to$ cause.
It should be noted that we consider both ``causality" and ``correlation" since 
it is well known that ``correlation" does not imply "causality" in recommendation, and ``correlation" means a kind of more general, undirected relationship, i.e., two items are purchased or consumed together.


\begin{figure}[htb]
    \centering
    \includegraphics[width=12.7cm]{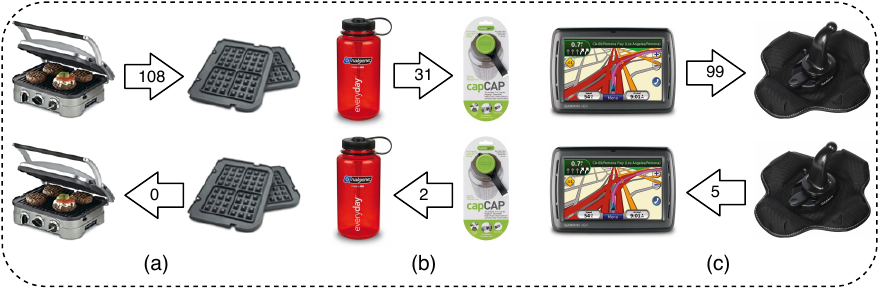}\vspace{-1mm}
    \caption{(a) griddlers and griddler waffle plates; (b) water bottles and capCAP; (c) GPS navigators and garmin portable friction mount.}\vspace{-1mm}
    \label{fig:amazonData}
\end{figure}

For example, we can easily observe this kind of directed cause$\rightarrow$effect relationship between items in real-world applications. Figure \ref{fig:amazonData} illustrates three examples mined from Amazon dataset\footnote{\url{jmcauley.ucsd.edu/data/amazon/links.html}.} \cite{he2016ups}. 
As can be viewed, the number of cases that firstly buy griddlers (or water bottles/GPS navigators) and then griddler waffle plates (or capCAP/garmin portable friction mount) is much higher than that of firstly buying griddler waffle plates (or capCAP/garmin portable friction mount) followed by griddlers (or water bottles/GPS navigators). The specific statistics are $108$ vs $0$, $31$ vs $2$, and $99$ vs $5$, respectively.
In summary, our motivation is not to capture all "causality" relationships but to explore the relationship between items from a "causality" perspective to achieve more effective recommendations.
\begin{figure}[ht]
    \centering
    \includegraphics[width=8cm]{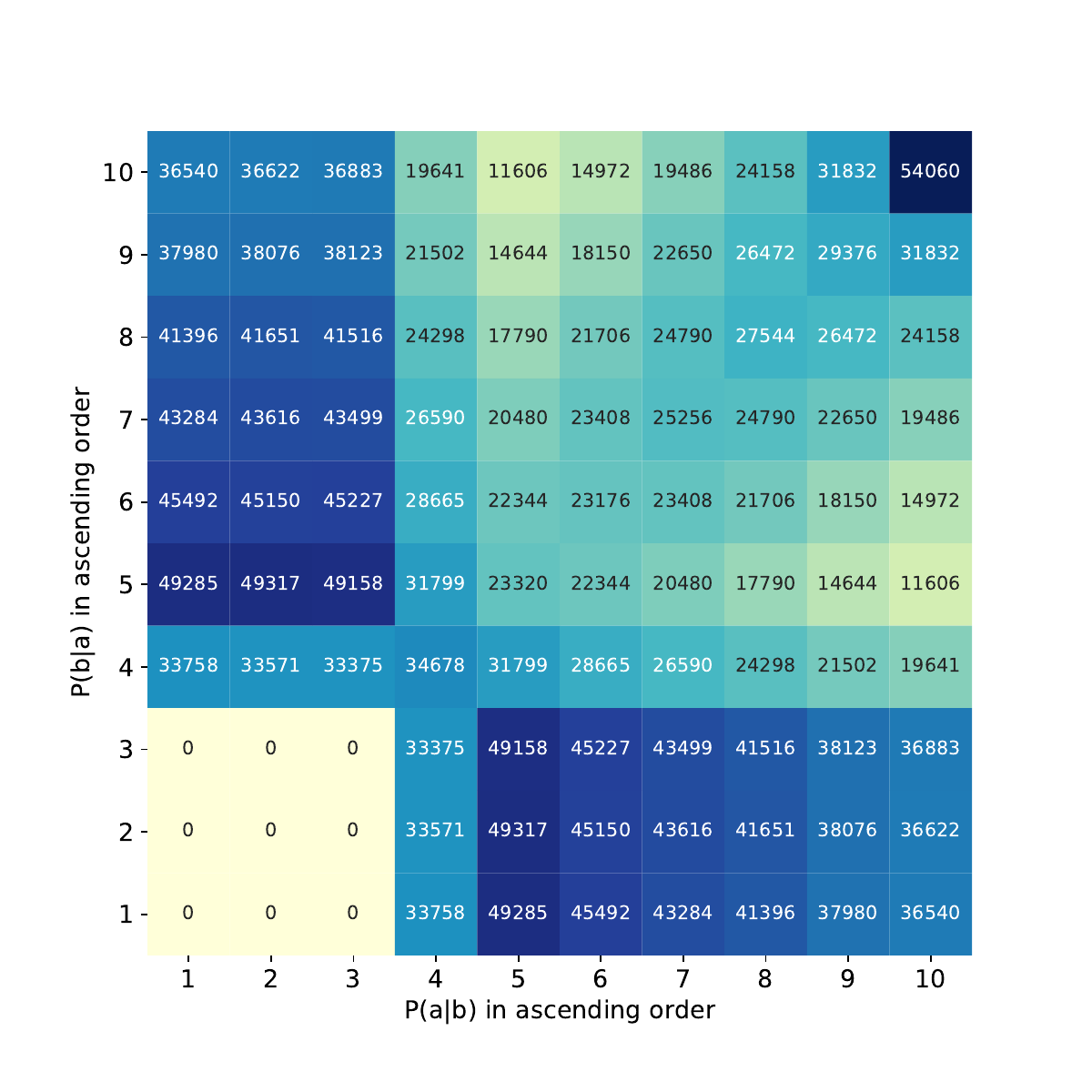}\vspace{-2mm}
    \caption{Causality statistics on Diginetica dataset.}\vspace{-2mm}
    \label{fig:digineticaStatistics}
\end{figure}
\begin{figure*}[htbp]
    \centering
    \hspace{-0.3in}
    \subfigure[Gowalla]{
    \includegraphics[width=0.5\linewidth]{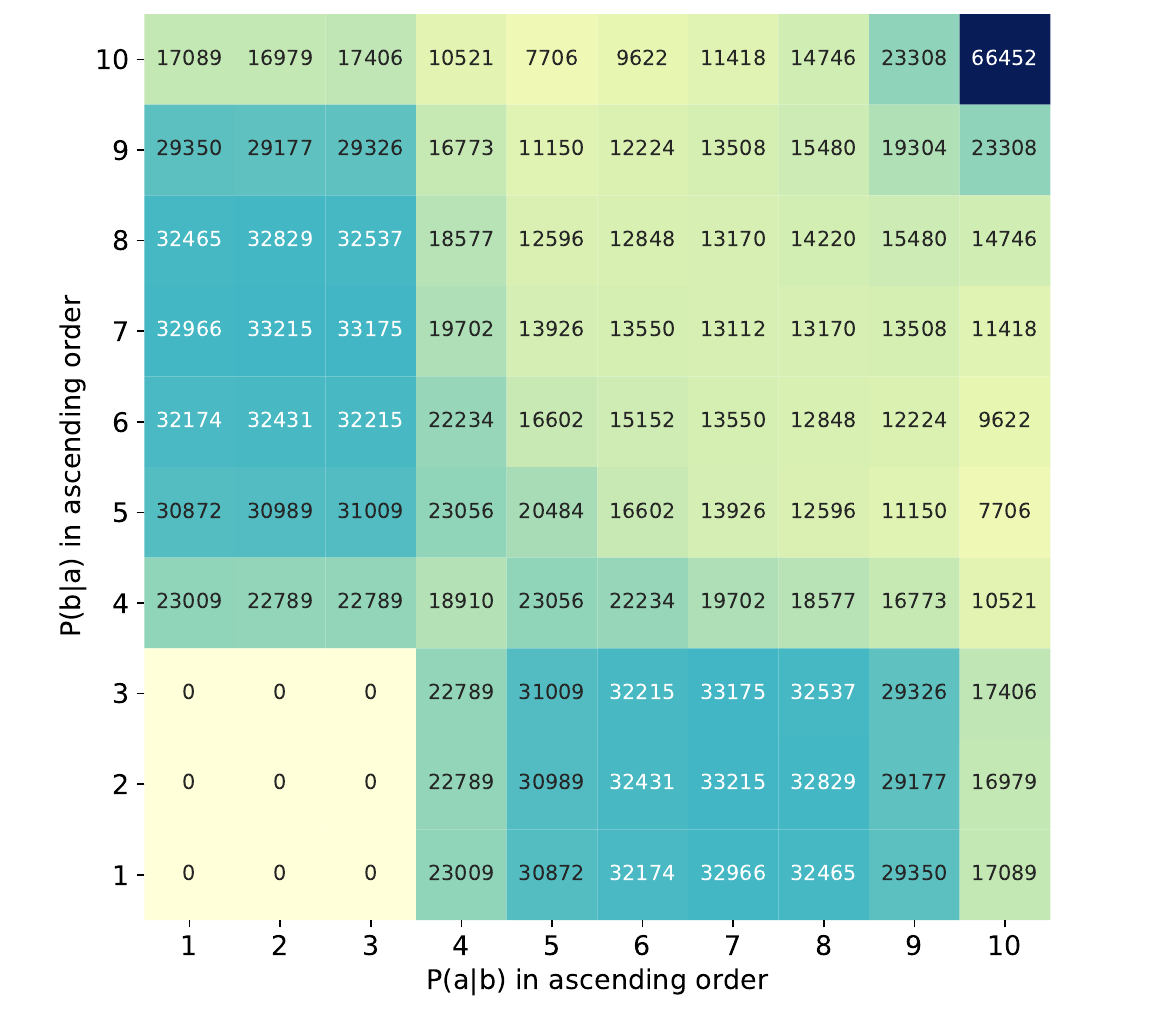}
    }
    \hspace{-0.3in}
    \subfigure[Amazon]{
    \includegraphics[width=0.5\linewidth]{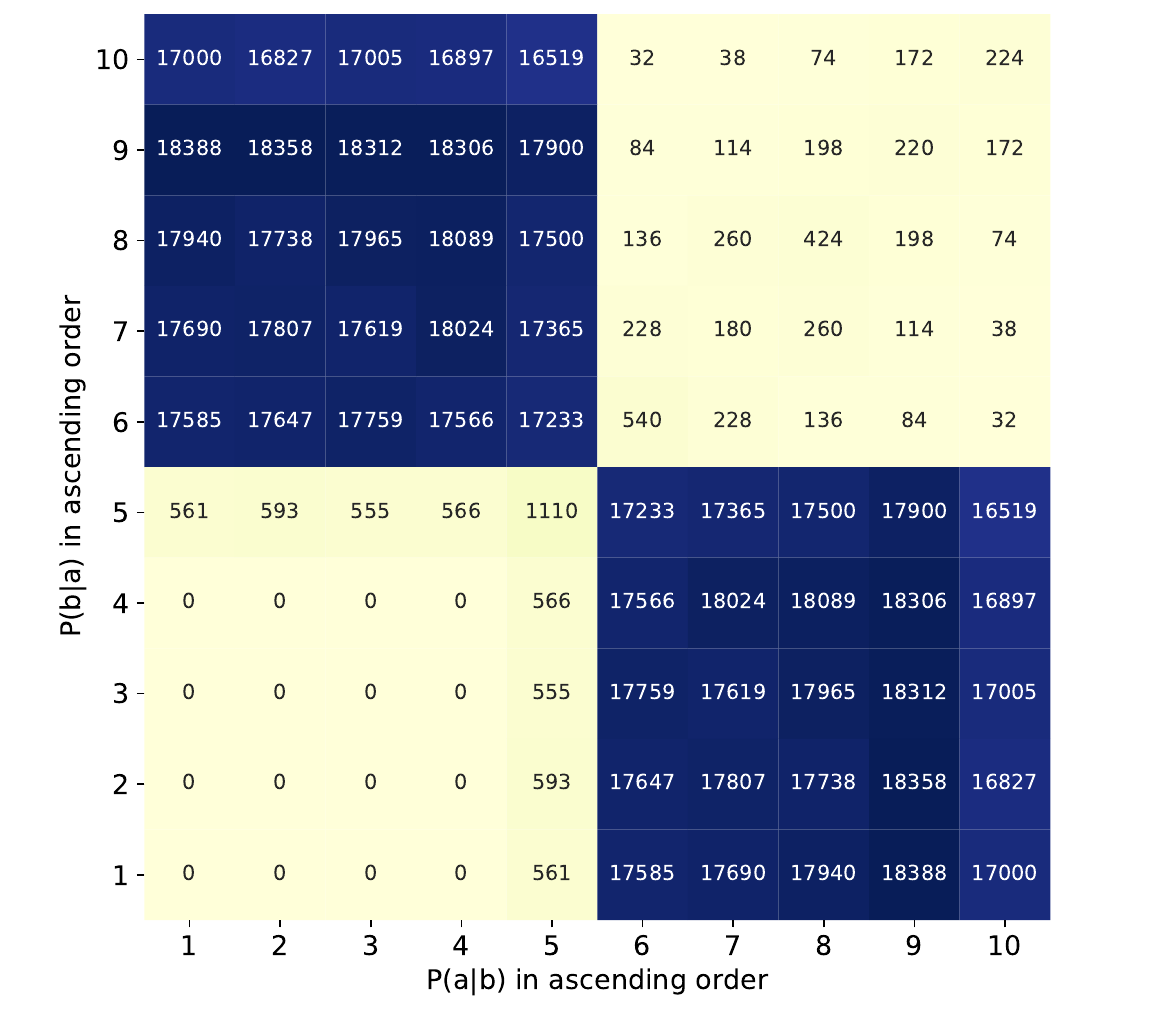}
    }
    \hspace{-0.3in}
   \caption{Causality statistics on two datasets.}\vspace{-2mm}
    \label{fig:causalityStatistics}
\end{figure*}

Then, we conduct a more comprehensive analysis to examine how to distinguish the causality and correlation between any two items on a typical session-aware dataset\footnote{\url{competitions.codalab.org/competitions/11161\#learn_the_details-data2}.}, i.e., Diginetica. In Figure \ref{fig:digineticaStatistics}, $p(a|b)$ ($a\neq b$) means, in a session, the probability of item $a$ that is interacted given the previously interacted item $b$. It is particularly calculated as $p(a|b)=\frac {\# (b \to a)}{\# b \to *}$, where $\# (b \to a)$ is the number of item $b$ interacted before item $a$ in the same session, and $\# b\to *$ is the frequency of item $b$ occurred before all other items. $p(b|a)$ is calculated in the same way. We further rank $p(a|b)$ in ascending order, and then divide all the item pair $(a,b)$ 
into ten groups ($\{1, 2,\cdots, 10\}$). Thus, each group has the same number of item pairs, 
and from group $1$ to $10$, $p(a|b)$ gets bigger. We deal with $p(b|a)$ similarly. Finally, we can place each item pair $(a,b)$ into a grid in terms of $p(a|b)$ and $p(b|a)$ as shown in Figure \ref{fig:digineticaStatistics}.
For example, $33,758$ in grid $(4,1)$ in Figure \ref{fig:digineticaStatistics} refers to the number of item pairs that $p(a|b)$ and $p(b|a)$ belong to the corresponding group $4$ and $1$, respectively. 
In this case, if $|p(a|b) - p(b|a)|\geq \epsilon$ ($\epsilon$ is a non-zero value), we consider that the relationship between items $a$ and $b$ is asymmetrical (i.e. sort of directed), and a larger $\epsilon$ indicates a more directed relationship between items $a$ and $b$. 
Accordingly, the relationship between item pairs in the bottom right (upper left) of Figure \ref{fig:digineticaStatistics} is more asymmetrical, revealing a higher possibility for being in the causal relations. 
Furthermore, the relationship in the upper right implies much stronger undirected correlation relationship (i.e., less possible causal relations), since there are quite lots of item pairs both from item $a$ to item $b$ and item $b$ to item $a$ simultaneously. Both of the directed and un-directed relations are quite prevalent (similar patterns can be viewed on Gowalla\footnote{\url{snap.stanford.edu/data/loc-Gowalla.html}.} and Amazon Home and Kitchen in Figure \ref{fig:causalityStatistics}), in this case, we should carefully consider both correlation and causality between items in item modeling for recommendation.

Therefore, considering the difference between the two types of item relationship, we propose a novel method called \underline{C}ausality and \underline{C}orrelation \underline{G}raph Modeling for Effective and Explainable \underline{S}ession-based \underline{R}ecommendation (CGSR) by particularly taking the causality and correlation relationship between items into consideration. Specifically, other than a correlation graph considering both first-order and three types of second-order relationship, we construct two graphs to capture causality relationship: a cause graph and an effect graph, which address the false causality by removing the impact of the common cause items given an item pair. Then, we design an end-to-end GNN-based model which takes the three graphs as input, outputs three kinds of item embeddings, and deploys attention mechanism to obtain corresponding session representations for final recommendation. Experimental results on 
three real-world datasets validate the superiority of our approach over the state-of-the-art. We further propose an explainable framework based on CGSR for SR, and conduct case studies on Amazon dataset to showcase that CGSR can also facilitate the explanation task in SR. 

The main contributions of this work are four-fold:
\begin{itemize}
\item To the best of our knowledge, we are the first to explore causality between items for SR, and propose CGSR to enhance the recommendation and explanation tasks in SR. 
\item We design an effective mechanism to capture possibly directed causality relationship between items. Particularly, we construct an effect graph and a cause graph on sessions which rule out the false causality by eliminating the impact of common cause items for every item pair.
\item We design a GNN-based method to combine causality and correlation graphs for effective recommendation. Regarding correlation graph, we consider both first-order and three types of second-order relationship. Exhaustive experiments on three real datasets verify the superiority of our approach over other baselines, and the validity of our designs.
\item We contribute to explainable SR by figuring out an explainable framework grounded on CGSR. Case studies on Amazon dataset demonstrate its usability and feasibility.
\end{itemize}

\section{Related Work}\label{sec:literature}
Our work is related to three primary tasks: session-based recommendation, item relation modeling in recommendation, and causal inference in recommendation. In the following subsections, we discuss each part to highlight our contributions over the related studies.

\subsection{Session-Based Recommendation}
Session-based recommendation (SR) predicts a user's next interested item(s) by deploying traditional approaches and deep learning (DL) techniques to process time-aware user-item interactions.
Traditional approaches \cite{shani2005mdp,liu2013personalized} apply machine learning (ML) techniques to capture item embedding in the session.
For example, 
FPMC \cite{rendle2010factorizing} applies matrix factorization (MF) and first-order Markov chains (MC) to address the sequential relationship among items. 
SEQ* \cite{le2016modeling} develops a hidden Markov model for sequences preference, which considers more factors, including two types of dynamic factors and contextual factors.
\citeauthor{wu2013personalized} \cite{wu2013personalized} propose a personalized Markov embedding (PME), which embeds songs and users into a Euclidean space, where the distances present the strengths of their relationships.
However, they fail to well capture the item relationships in relatively longer session sequences. It is worth mentioning that MC-based methods also define the directed relationship between items, which indicates that two items incline to be dependent with each other. However, this relationship is not equal to causality. Besides, they possibly ignore the correlation relationship.

On the contrary, DL methods \cite{xia2021self,gwadabe2022improving} are capable of dealing with a much longer sequence than traditional models.
GRU4Rec \cite{hidasi2015session} firstly applies recurrent neural network (i.e., a multi-layer gate recurrent unit) to process session data.
Later, there are a lot of variants with regard to GRU4Rec. For instance, 
HRNNs \cite{quadrana2017personalizing} extends it to the hierarchical form which simultaneously considers both short-term and long-term preferences with two GRU constructs, i.e., the session-level GRU (GRUses) and the user-level GRU (GRUusr). 
\citeauthor{donkers2017sequential} \cite{donkers2017sequential} model the temporal dynamics of consumption sequences based on the gated RNN and explicitly represent the individual user in a gated architecture.
NARM \cite{li2017neural} combines GRUs and vanilla attention mechanism to better extract main purpose from the current session, which can effectively eliminate noise from unintended behaviors.
However, these methods mainly address behavior dependency in a session, but cannot directly capture item relationship across different sessions. Besides, they ignore to distinguish causality relationship from correlation relationship between items.

With the rapid development of graph neural networks (GNN) in recent years, we have witnessed its great success in many downstream tasks, e.g., node classification and recommender systems.
Therefore, some studies have started to deploy GNNs for session-based recommendation, and obtained encouraging results \cite{pan2020star,qiu2020exploiting,qiu2020gag}.
For example, SR-GNN \cite{wu2019session} firstly combines different sessions into session graphs, and then uses GNN \cite{li2015gated} to learn representation of each item and finally obtain session representations through attention mechanism.
Experimental results on Yoochoose and Diginetica verify that it can obtain better performance than both RNN-based models (e.g., GRU4Rec) and attention based models (e.g., STAMP \cite{liu2018stamp}).
Later, quite a few variants of SR-GNN \cite{xu2019graph, chen2020handling,wang2020global} have been proposed. For example,
\citeauthor{xu2019graph} \cite{xu2019graph} introduce a novel graph contextual self-attention model
based on the graph neural network called GC-SAN, which obtains local graph structured dependencies of separated session sequences and models contextualized non-local representations.
LESSR \cite{chen2020handling} further reduces information loss by proposing an edge-order preserving aggregation layer and a shortcut graph attention layer.
Moreover, hypergraph networks are also applied in session-based recommendation. For instance, SHARE \cite{wang2021session} proposes the hypergraph structure and hypergraph attention networks, which exploit the relationship among items within various contextual windows.
However, as have been discussed, all GNN-based methods have not directly considered the directed causality relationship between items, which might lead to incorrect recommendations.

\subsection{Relation Modeling in Recommendation}
Quite a few studies target to explore various relationship between items using side information like textual and visual information \cite{mcauley2015image,zhang2018quality}. For example, 
Sceptre \cite{mcauley2015inferring} casts the item relations identification problem as a supervised link prediction task, and predicts substitutable and complementary items by learning latent topics from textual information.
\citeauthor{zhang2018quality} \cite{zhang2018quality} propose a neural complementary recommender Encore which can jointly learn complementary item relationship and user preferences through Bayesian inference. 
However, these studies aim to design specific models for identifying the well-studied relationship (e.g., substitute and complementary) between items from side information other than interaction data. On the contrary, in our study, relying on the sequential interaction data in sessions, we try to identify the directed causality relationship between items for effective recommendation. Our causality relationship is expected to complement the identified item relationship widely discussed in the business area.
\subsection{Causal Inference in Recommendation}
There are also some studies that have explored the causal inference for recommendation \cite{liang2016causal, schnabel2016recommendations, wang2018deconfounded,wang2019doubly}. For example, 
\citeauthor{bonner2018causal} \cite{bonner2018causal} optimize the causal recommendation outcomes via user implicit feedback based on the factorizing matrices, which presents that the objective of causal recommendations is equal to factorizing a matrix of user responses.
\citeauthor{wang2020causal} \cite{wang2020causal} consider that the core of recommender system is to address a causal inference question 
by solving two problems: which items the users decide to interact with, and how the users rate them.
\citeauthor{qiu2021causalrec} \cite{qiu2021causalrec} proposes a deconfounded recommender, which utilizes Poisson factorization to infer confounders in treatment assignments.
CauSeR \cite{gupta2021causer} provides a more holistic causal view of item popularity related biases at two stages, i.e., data generation and training stages. 
CR-VAR \cite{xu2021learning} designs a post-hoc causal explanation for the black-box sequential recommendation methods, where the causal explanations are obtained through a perturbation model and a causal rule learning model.

We can see that this line of research is more related to unbiased recommendation and tries to reason about personalized user preference well, which is quite different from our research scenario. In our study, we try to explore the directed causality relationship between items to facilitate session-based recommendation.

\section{Graph Construction}\label{sec:prelim}
In this section, we firstly formally define the research problem, and then present how to construct causality and correlation graphs from sessions, as well as the edge weights in great details. 

\subsection{Problem Statement}
In session-based recommendation, let $\mathcal{V}=\{v_1,v_2,\cdots, v_N\}$ denote the set of items ($|\mathcal{V}|=N$). An anonymous session $S$ can be represented by $S=\{v_1^S,v_2^S,\cdots,v_l^S\}$, where its length equals to $l$ and $v_i^S\in \mathcal{V}$ means the $i$-th interacted item within $S$. There are $M$ sessions in total.
Given session $S$, the goal of session-based recommendation is to predict the next item (i.e., the $l+1$-th) that will be purchased. Therefore, in our CGSR, we strive to firstly construct graphs (i.e., causality and correlation graphs) from training sessions, and then learn effective representation of session $S$. Thirdly, we generate recommendation score $\widehat{y}_j$ for each candidate item $v_j\in \mathcal{V}$, and finally recommend top $K$ items with the highest recommendation scores.

Next, we will elaborate causality and correlation graphs construction in detail, respectively.
\subsection{Constructing Causality Graphs}
While constructing causality graphs from sessions, we aim to maximally identify the truly directed causality relationship meanwhile ruling out the false ones (i.e., the noisy information).
To fulfill the goal, as shown in Figure \ref{fig:CGSRmodel}(a), we firstly build a \emph{session graph} from sessions, on the basis of which, we then construct an \emph{effect graph} and a \emph{cause graph} by removing the impact of noisy information. Noted that, in causality graphs, we only consider first-order relationship since we strive to directly extract the most probability causal relations between items given historical data and high-order ones might involve more noise. Besides, GNN model is supposed to automatically capture high-order relations.

\begin{figure}[htbp]
    \centering
    \includegraphics[width=12.7cm]{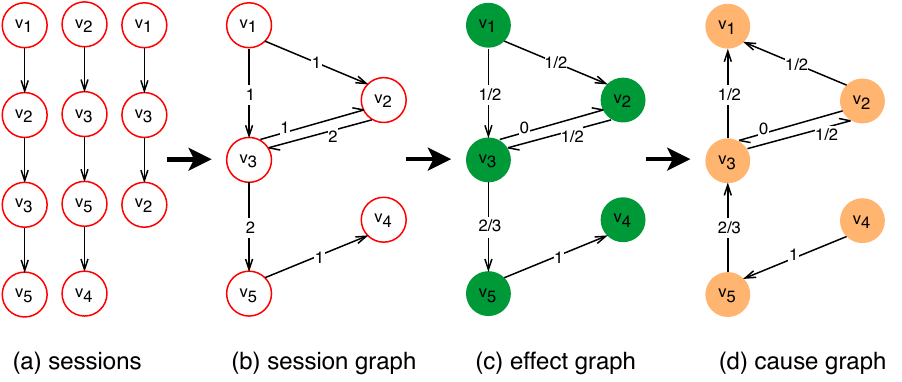}\vspace{-1mm}
    \caption{The process of constructing causality graphs (i.e., an effect graph and a cause graph).}\vspace{-1mm}
    \label{fig:causalityGraph}
\end{figure}

\textbf{Session graph construction.} Without the loss of generality, 
let $\mathcal{G}^s=(\mathcal{V}^s,\mathcal{E}^s)$ be the correspondingly directed session graph, where $\mathcal{V}^s$ indicates the node set which is identical to the item set in all sessions, and $\mathcal{E}^s$ denotes the edge set. An edge $E^s_{i,j}\in \mathcal{E}^s$ refers to that in a session, item $v_i$ is firstly interacted followed by item $v_j$, and $w^s_{i,j}$ is the corresponding occurred frequency to denote the weight of edge $E^s_{i,j}$.
For example, in Figure \ref{fig:causalityGraph}, these three sessions in Figure \ref{fig:causalityGraph} (a) can form a directed session graph in Figure \ref{fig:causalityGraph} (b).

\textbf{Causality graphs construction}. To better represent the causality relationship, we construct two graphs: an effect graph ($\mathcal{G}^e$) and a cause graph ($\mathcal{G}^{c})$ on the basis of the session graph. Specifically, in \emph{effect graph}, we want to learn representation of an item where the item is playing the \emph{effect} role in the directed cause-effect relationship between two items. Consequently, each item representation is expected to integrate the information of the adjacent cause items by information propagation via GNN. In contrast, the \emph{cause graph} is to capture the item information where the item is playing the \emph{cause} role in the directed relationship. We clarify the following two issues: (1) the node and edge sets of the effect graph are initially the same as those of session graph, mainly because the session graph models the temporally directed relationship between items, basically indicating all the possible causality relationship from sessions. However, as $\mathcal{G}^s$ may involve false causality relationship (elaborated later), we will eliminate 
its impact by justifying the corresponding weight of each edge; (2) the cause graph is quite similar to the effect graph, except that the direction of each edge is opposite to that of effect graph, considering that in cause graph, we want to explore the part of item information leading to the purchase of other items.

Towards the first issue, we elaborate \emph{how to get rid of the impact of false causality relationship}. We mainly adopt the idea of bias of confounders in causal inference \cite{keith2020text}, where confounders denote variables that influence both independent variables and the dependent variable. In causal inference, estimating the effect of an independent variable on the dependent variable without accounting for confounders could result in strongly biased estimates and thus invalid causal conclusions. Therefore, we adapt this idea into our scenario. Let us first view an example in session-based recommendation. There are three sessions: $S_1$ [``iPhone'', ``charging line'', ``charger'', ``phone shell''], $S_2$ [``iPhone'', ``charger'', ``charging line''], and $S_3$ [``iPhone'', ``charging line'', ``phone shell'']. Based on the three sessions (training data), a phone shell will be recommended given session [``charging line'', ``charger''], which is quite odd since few users will purchase a phone shell if having previously bought a charging line and charger. By examining the data, intuitively, we see that ``owing an iPhone'' is a \emph{common cause} to also purchase a charging line, charger and phone shell. In this case, such odd recommendation is probably
induced by the false causality relationship led by the common cause (i.e., ``iPhone'') in training data.
That is, purchasing ``iPhone'' leads to the purchase of ``charger'' and ``phone shell'', instead of purchasing ``charger'' causing the purchase of ``phone shell''.

Therefore, to overcome this issue, we appropriately calculate the weight of each edge in $\mathcal{G}^e$ by taking the common cause of two items into consideration. Specifically, for each item pair $v_i$ and $v_j$, we firstly identify every common cause $v_k\in I^s_i\cap I^s_j$ ($I^s_*$ is the node set directed into node $v_*$ in $\mathcal{G}^s$). For example, in Figure \ref{fig:causalityGraph} (b),
$v_1$ is the common cause to $v_2$ and $v_3$. Secondly, to identify the true causality strength of $v_i\to v_j$ in $\mathcal{G}^e$, we eliminate the impact of every common cause $v_k$ to calculate the weight of $E^e_{i,i}$, $w^e_{i,j}$:
\begin{equation}
\small
w^e_{i,j}=\frac{w^s_{ij}-\sum\limits_{v_k\in I^s_i\cap I^s_j}\#[v_k,v_i,v_j]}{\#[v_i,v_*]}\label{eq:effectWeight}
\end{equation}
where $\#[v_k,v_i,v_j]$ and $\#[v_i,v_*]$ are the number of sequence [$v_k$, $v_i$, $v_j$] and the number of sequence [$v_i$, $v_*$] ($v_*\in V$) existing in all sessions (training data), respectively.
For example, the weights of edges in Figure \ref{fig:causalityGraph} (c) are computed as: $w_{1,2}=(1-0)/2$; $w_{1,3}=(1-0)/2$; $w_{2,3}=(2-1)/2$; $w_{3,2}=(1-1)/2$; $w_{3,5}=(2-0)/3$; $w_{5,4}=(1-0)/1$. After obtaining the effect graph $\mathcal{G}^e$, we reverse its directions of edges to get the cause graph $\mathcal{G}^{c}$.


\subsection{Constructing Correlation Graph}

\begin{figure}[htbp]
    \centering
    \includegraphics[width=10cm]{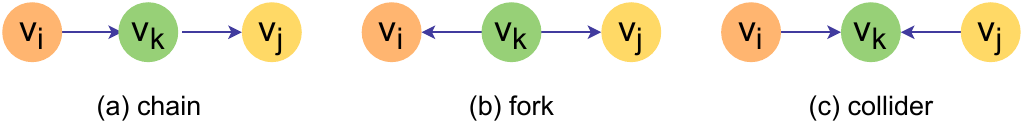}\vspace{-1mm}
    \caption{Three types of second-order relationship.}\vspace{-1mm}
    \label{fig:secondRelationships}
\end{figure}

Besides the directed causality relationship between items, we also consider the undirected correlation relationship, whose effectiveness in SR has been validated in previous GNN-based studies \cite{wu2019session,wang2020global}. 
While constructing the correlation graph, apart from the generally adopted \emph{first-order} relationship (neighbor in sequence), we additionally consider the \emph{second-order} relationship (neighbor of neighbor), by following the study of \cite{wang2020global}. Noted that differing from \cite{wang2020global}, we distinguish three kinds of second-order relationship (i.e., \emph{chain}, \emph{fork}, and \emph{collider}, see Figure \ref{fig:secondRelationships}) for better exploring the correlation relationship for effective recommendation. 

\begin{figure}[htb]
    \centering
    \includegraphics[width=12.7cm]{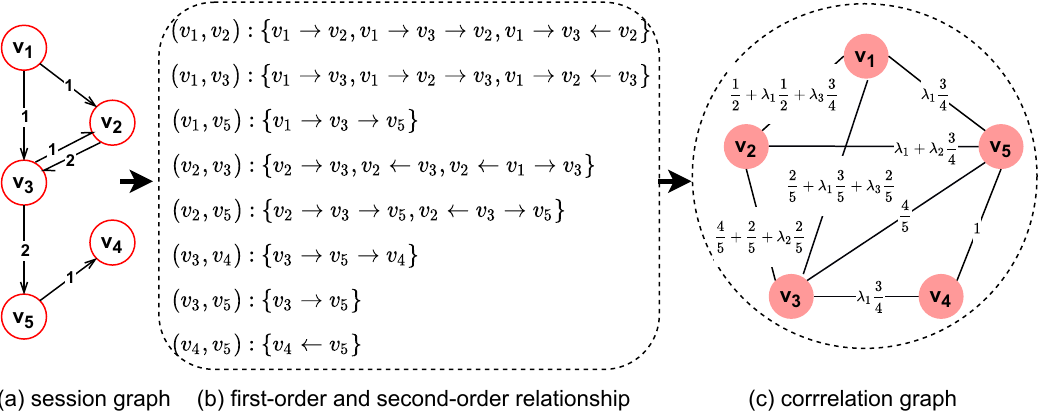}\vspace{-1mm}
    \caption{The process of constructing correlation graph.}\vspace{-1mm}
    \label{fig:correlationGraphs}
\end{figure}

Particularly, as shown in Figure \ref{fig:correlationGraphs}, based on the session graph $\mathcal{G}^s$ have been constructed, we calculate the weight of each possible edge by considering the first-order and second-order neighbors.

\textbf{For the first-order relationship}, namely ($v_i,v_j$) in session graph $\mathcal{G}^s$, considering $\mathcal{G}^s$ is directed whilst correlation graph $\mathcal{G}^r$ is undirected, the first-order weight of each edge,  $w^{r,1}_{i,j}$, is computed as:
\begin{equation}
\small
w^{r,1}_{i,j}=\underset{\text{impact of $v_i\to v_j$}}{\underbrace{\frac{2*w^s_{i,j}}{\sum\limits_{v_k\in O^s_i} w^s_{i,k} + \sum\limits_{v_k\in I^s_j} w^s_{k,j}}}}+\underset{\text{impact of $v_j\to v_i$}}{\underbrace{\frac{ 2*w^s_{j,i}}{\sum\limits_{v_k\in O^s_j} w^s_{j,k} + \sum\limits_{v_k\in I^s_i} w^s_{k,i}}}}\label{eq:firstWeight}
\end{equation}
where $O^s_*$ and $I^s_*$ refer to the node (item) set directed out and into item $v_*$ in $\mathcal{G}^s$ respectively.

\textbf{For the second-order relationship}, we first extract all the possible second-order relationship for each item pair $v_i$ and $v_j$ based on $\mathcal{G}^s$. That is, if $v_k$ is the neighbor of both $v_i$ and $v_j$, we say that there is a second-order correlation between $v_i$ and $v_j$. In particular, with the consideration on the link directions among the corresponding three items, we identify three types of second-order relationship, intuitively denoted as \emph{chain}, \emph{fork} and \emph{collider} (see Figure \ref{fig:secondRelationships}). We treat the three types separately mainly because they act differently on casting correlation relationship between items \cite{taskar2003link}, which are modeled as different weighting factors in our study (see $\lambda_1$, $\lambda_2$ and $\lambda_3$ in Equation \ref{eq:secondorderWeight}).
Thus, the weight of link ($v_i$, $v_j$) in terms of second-order correlation, $w^{r,2}_{i,j}$, is thus computed as:
\begin{equation}
\small
\begin{aligned}
w^{r,2}_{i,j}=&\underset{\text{chain}}{\underbrace{\lambda_1\frac{\sum\limits_{v_k\in O^s_i\cap I^s_j}\left(w^s_{i,k}+w^s_{k,j}\right)}{\sum\limits_{v_k\in O^s_i} w^s_{i,k} + \sum\limits_{v_k\in I^s_j} w^s_{k,j}}+\lambda_1\frac{\sum\limits_{v_k\in I^s_i\cap O^s_j}\left(w^s_{j,k}+w^s_{k,i}\right)}{\sum\limits_{v_k\in O^s_j} w^s_{j,k} + \sum\limits_{v_k\in I^s_i} w^s_{k,i}}}} \\
+&\underset{\text{fork}}{\underbrace{\lambda_2\frac{\sum\limits_{v_k\in I^s_i\cap I^s_j}\left(w^s_{k,i}+w^s_{k,j}\right)}{\sum\limits_{v_k\in I^s_i} w^s_{k,i} + \sum\limits_{v_k\in I^s_j} w^s_{k,j}  }}}+\underset{\text{collider}}{\underbrace{
\lambda_3\frac{\sum\limits_{v_k\in O^s_i\cap O^s_j}\left(w^s_{i,k}+w^s_{j,k}\right)}{\sum\limits_{v_k\in O^s_i} w^s_{i,k} + \sum\limits_{v_k\in O^s_j} w^s_{j,k}}}}
\end{aligned}\label{eq:secondorderWeight}
\end{equation}
where $\lambda_1$, $\lambda_2$ and $\lambda_3$ are trainable parameters to balance the impact of the three second-order relationship types.

Finally, we get the weight of every edge $E^r_{i,j}$ in correlation graph $\mathcal{G}^r$, $w^r_{i,j}$, by adding the first-order and second-order weights, namely,  $w^r_{i,j}=w^{r,1}_{i,j}+w^{r,2}_{i,j}$, as depicted in Equation \ref{eq:firstWeight} and \ref{eq:secondorderWeight}. Noted that compared with $\mathcal{G}^s$, $\mathcal{G}^r$ can obtain new edges by the three types of second-order correlations, e.g., $E^r_{1,5}$, $E^r_{2,5}$, $E^r_{3,4}$ in Figure \ref{fig:correlationGraphs} (c).

\section{The CGSR Model}\label{sec:model}

\begin{figure*}[htb]
    \centering
    \includegraphics[width=12.7cm]{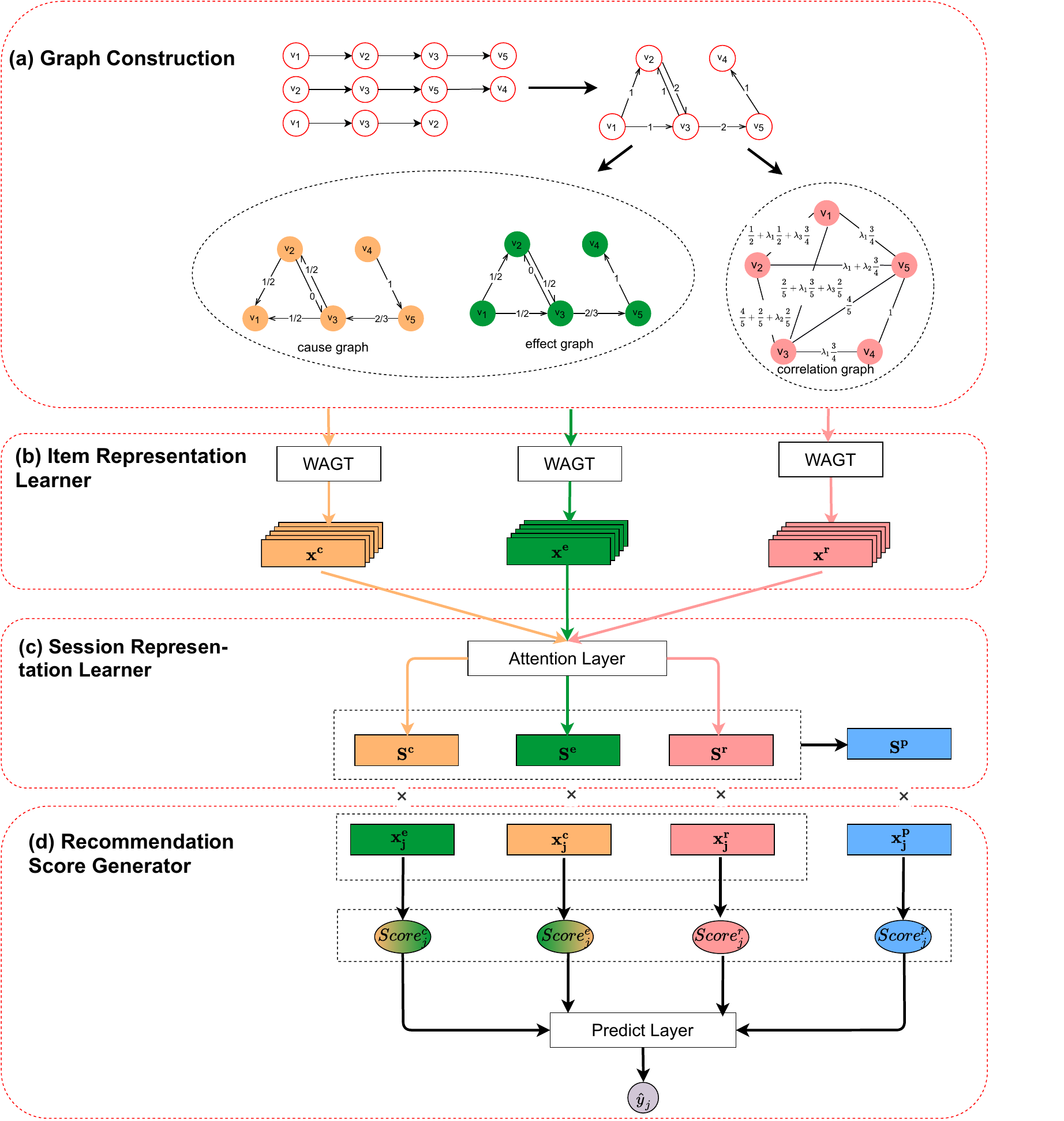}\vspace{-2mm}
    \caption{The overview of our proposed CGSR model.
    }\vspace{-2mm}
    \label{fig:CGSRmodel}
\end{figure*}

In this section, we present our proposed CGSR model detailedly. Figure \ref{fig:CGSRmodel} outlines the overview of CGSR, which consists of four components: (a) \emph{Graph Construction}, (b) \emph{Item Representation Learner}, (c) \emph{Session Representation Learner} and (d) \emph{Recommendation Score Generator}. In particular, we firstly build three types of graphs (i.e., effect graph $\mathcal{G}^e$, cause graph $\mathcal{G}^c$ and correlation graph $\mathcal{G}^r$) in \emph{Graph Construction} as introduced in Section \ref{sec:prelim}. \emph{Item representation Learner} deploys a weighted graph attention network (WGAT) on each of the three graphs to obtain 
a representation for each item, respectively. That is, for each item $v_i$, we obtain three types of representations, namely $\mathbf{x_i}^e$, $\mathbf{x_i}^c$, and $\mathbf{x_i}^r$, given $\mathcal{G}^e$, $\mathcal{G}^c$ and $\mathcal{G}^r$. \emph{Session Representation Learner} uses an Attention Layer to aggregate each type of learned item representation in session sequence $S$ to obtain session representation $\mathbf{S}^e$, $\mathbf{S}^c$ and $\mathbf{S}^r$, respectively. We also get session representation $\mathbf{S}^p$ by averaging the three types of session representation. \emph{Recommendation Score Generator} strives to calculate the recommendation score, $\hat{y}_j$ of each candidate item $v_j\in \mathcal{V}$ on the basis of the learned item and session representations.
Next, we elaborate the last three components of CGSR.

\subsection{Item Representation Learner}
\label{subsec:itemRepresentation}

Here, we aim to learn item embedding on built graphs $\mathcal{G}^c$, $\mathcal{G}^e$ and $\mathcal{G}^r$. Considering that the three graphs are either weighted (correlation graph) or simultaneously directed and weighted (cause and effect graphs), we thus adopt weighted graph attention network (WGAT) \cite{qiu2019rethinking} to obtain item representations.
Specifically, we denote $\mathbf{X}^0 \in \mathbb{R}^{N\times d_0}$ as initial embedding matrix of item set (for item $v_i \in \mathcal{V}$, initial item embedding $\mathbf{x_i}^0=\mathbf{X}^0_{i,:}\in\mathbb{R}^{d_0}$):
\begin{equation}
\small
\mathbf{X}^0=\text{nn.Embedding}(N, d_0)\label{eq:initialnnEmbedding}
\end{equation}
where nn.Embedding() is a function in PyTorch, which randomly initializes a vector following a normal distribution $N(0,1)$.

Then, taking cause graph $\mathcal{G}^c$ as an example, in WGAT, self-attention mechanism is deployed to aggregate information from each node (item) $v_i$'s directed-into neighbors. As in our scenario, a session is normally not very long, all first-order neighbors are considered. Thus, the importance between $v_i$ and its neighbor $v_j$ ($v_j\in I^c_i$, and $I^c_i$ is the node set directed into $v_i$ in $\mathcal{G}^c$) , $e^c_{i,j}$ (i.e., self-attention coefficient) is computed as:
\begin{equation}
\small
e^c_{ij}=\sigma(\mathbf{W}_{c,2}^T*[\mathbf{W}_{c,1}\mathbf{x}_i^0; \mathbf{W}_{c,1}\mathbf{x}_j^0; w^c_{j,i}]) \label{eq:selfAttentionWeight}
\end{equation}
where $\sigma(.)$ is the Leaky ReLU function, $\mathbf{W}_{c,1}\in \mathbb{R}^{d\times d_0}$ and $\mathbf{W}_{c,2}\in \mathbb{R}^{2d+1}$ are trainable parameters, and $w^c_{j,i}$ is the weight of link $v_j\to v_i$.
Softmax function is further adopted to normalize the $e^c_{ij}$:
\begin{equation}\small
\alpha^c_{ij}=\text{softmax}(e^c_{ij})=\frac {exp(e^c_{ij})}{\sum\limits_{v_k\in I^c_i} exp(e^c_{ik})}\label{eq:normalizedAttentionValue}
\end{equation}

Third, the information from neighbors are weighted to get item $v_i$'s embedding, $\mathbf{x_i^c}\in\mathbb{R}^{d}$:
\begin{equation}\small
\mathbf{x}_{i}^c=\sigma (\sum\limits_{j\in I^c_i} \alpha^c_{ij}\mathbf{W}_{c,3}\mathbf{x}_j^0) \label{eq:oneHead}
\end{equation}
where $\mathbf{W}_{c,3}\in\mathbb{R}^{d\times d_0}$ is a trainable parameter matrix, and $\sigma(.)$ is the Leaky ReLU function.

In multi-head attention mechanism of WGAT, we average obtained embedding from $K_m$ heads to output final item embedding:
\begin{equation}\small
\mathbf{x_i^c}=\frac{1}{K_m}*\sum\limits_{k=1}^{K_m}\mathbf{x_{i,k}^c}\label{eq:finalItemEmbedding}
\end{equation}
where $\mathbf{x}_{i,k}^c$ is $v_i$'s representation output by the $k$-th head using Equations \ref{eq:selfAttentionWeight}-\ref{eq:oneHead}.

Similarly, we get the embedding of $v_i$, $\mathbf{x}_i^e$ and $\mathbf{x}_i^r$, from effect graph $\mathcal{G}^e$ and correlation graph $\mathcal{G}^r$ respectively.

\subsection{Session Representation Learner}
\label{subsec:sessionRepresentaion}
In \emph{Item Representation Leaner}, we obtain three embedding for each item $v_i$ in terms of three graphs. Specifically, $\mathbf{x}_i^c$ considers information from $v_i$'s effect neighbors, whilst $\mathbf{x}_i^e$ and $\mathbf{x}_i^r$ involve information from its cause and correlated neighbors, respectively.

In contrast to previous studies \cite{wang2020global} fusing embedding from various sources to generate session representation, in CGSR, \emph{Session Representation Learner} treats each type of item representation separately. That is, it learns a specific session representation given each type of item embedding, for the purpose of obtaining a rather pure session representation for each relationship instead of losing information by aggregating them too early.

Specifically, for instance, in terms of cause graph $\mathcal{G}^c$, to generate the representation of session $S$ ($S=\{v_1^S,\cdots,v_l^S\}$), $\mathbf{S}^c$, using attention mechanism, we first compute the weighted factor $\alpha^c_k$ depicting the importance of $k$-th item ($v_k^S$) to $l$-th item (last item, $v_l^S$) in $S$:
\begin{equation}\small
\alpha^c_k=\mathbf{q}_c^{T} \sigma (\mathbf{W}_{c,4} \mathbf{x}_{l,S}^c + \mathbf{W_{c,5}} \mathbf{x}_{k,S}^c +\mathbf{b}_c)\label{eq:itemImportanceToSession}
\end{equation}
where $\mathbf{q}_c\in\mathbb{R}^d$ denotes the weighting vector. $\mathbf{W}_{c,4}$, $\mathbf{W}_{c,5} \in \mathbb{R}^{d\times d}$ are the weighting matrices. $\mathbf{b}_c\in \mathbb{R}^d$ is the bias vector. $\sigma(.)$ is the sigmoid function. $\mathbf{x}_{l,S}^c$ and $\mathbf{x}_{k,S}^c$ denote the embedding of $l$-th and $k$-th items given cause graph respectively. We thus use weighted factors to aggregate all item information for session representation:
\begin{equation}\small
\mathbf{S}_g^c=\sum\limits_{k=1}^l \alpha_k^c \mathbf{x}_{k,S}^c \label{eq:globalSession}
\end{equation}
Following the previous studies, we also particularly consider the information of the most recent behavior in the session (i.e., $v_l^S$), namely, to get the concatenation of $\mathbf{S}_g^c$ and $\mathbf{x}_{l,S}^c$. We further project the concatenation to get the final session representation via:
\begin{equation}\small
\mathbf{S}^c=\mathbf{W}_{c,6} [\mathbf{x}_{l,S}^c; \mathbf{S}_g^c] \label{eq:sessionRepresentation}
\end{equation}
where $\mathbf{W}_{c,6}\in \mathbb{R}^{d\times 2d}$ is the projecting matrix.

Similarly, we can learn the session representation on effect and correlation graphs, $\mathbf{S}^e$ and $\mathbf{S}^r$, respectively. We also generate a session representation ($\mathbf{S}^p$, referred as preference-related session representation) by fusing the three types of session representation using mean operator, and further project it into a new latent space:
\begin{equation}\small
\mathbf{S}^p=\mathbf{W}_{7}*\text{mean} (\mathbf{S}^c, \mathbf{S}^e, \mathbf{S}^r)\label{eq:fusedSessionRepresentation}
\end{equation}
where $\mathbf{W}_{7}\in\mathbb{R}^{d\times d}$ is the projecting matrix. $\text{mean}{(*)}$ function outputs the corresponding average value.

In summary, \emph{Session Representation Leaner} outputs four session representations of session $S$, i.e., $\mathbf{S}^c$, $\mathbf{S}^e$, $\mathbf{S}^r$, and $\mathbf{S}^p$.


\subsection{Recommendation Score Generator}
\label{subsec:recommendationScore}
Given the learned session representations of session $S$, for each candidate item $v_j$, \emph{Recommendation Score Generator} will output its recommendation score. Particularly, the final score is three-fold: (1) causality score; (2) correlation score; and (3) preference score.

\emph{Causality score} strives to maximize item transition from cause $\rightarrow$ effect whilst simultaneously minimize that from effect $\to$ cause. Accordingly, causality score of item $v_j$, $Score_j^{ca}$, is calculated as:
\begin{equation}\small
Score_j^{ca}=Score_j^{c}-Score_j^{e}=(\mathbf{S}^c)^T \mathbf{x}_j^e-\gamma_1 (\mathbf{S}^e)^T \mathbf{x}_j^c \label{eq:causalityscore}
\end{equation}
where $\gamma_1$ is a trainable parameter to balance the two components.

\emph{Correlation score} and \emph{preference score} of $v_j$, $Score_j^r$ and $Score_j^p$, are defined as:.
\begin{equation}\small
\begin{aligned}
Score_j^r=(\mathbf{S}^r)^T \mathbf{x}_j^r 
; \ \ \  Score_j^p=(\mathbf{S}^p)^T \mathbf{x}_j^p
\end{aligned}\label{eq:preferenceScore}
\end{equation}
where $\mathbf{x}_j^p=\text{mean}(\mathbf{x}_j^c,\mathbf{x}_j^e,\mathbf{x}_j^r)$.

Finally, we compute the overall score of candidate item $v_j$ ($Score_j$):
\begin{equation}\small
Score_j=Score_j^p + \gamma_2  Score_j^{ca} + \gamma_3  Score_j^r  \label{eq:Score}
\end{equation}
where $\gamma_2$ and $\gamma_3$ are trainable parameters. Softmax function is further deployed to obtain the final recommendation score $\hat{y}_j$: 
\begin{equation}\small
\hat{y}_j=\text{softmax}(Score_j)=\frac{exp(Score_j)}{\sum\limits_{v_k\in\mathcal{V}}exp(Score_k)}  \label{eq:recommendationScore}
\end{equation}

We adopt cross-entropy loss to train the CGSR model:
\begin{equation}\small
L=-\sum_{j=1}^{N} y_j \log(\hat{y}_j)+(1-y_j)\log (1-\hat{y}_j) \label{eq:loss}
\end{equation}
where $y_j$ is the ground-truth of item $v_j$ (1 or 0, $1$ indicates that $v_j$ is the next interacted item and $0$ means that $v_j$ is not).

\section{Empirical Evaluations}
\label{sec:experiments}

In this section, we conduct extensive experiments on three datasets to validate the effectiveness of our proposed CGSR, with the goal of answering four specific research questions (RQs):
\begin{itemize}
\item \textbf{RQ1}: How does CGSR perform compared to other state-of-the-art approaches?
\item \textbf{RQ2}: How do different components of CGSR (e.g., causality graphs) contribute to the recommendation performance? 
\item \textbf{RQ3}: How do different hyper-parameters affect the performance of CGSR?
\item \textbf{RQ4}: How does CGSR facilitate the explanation task of session-based recommendation?

\end{itemize}
\subsection{Experimental Setup}

\subsubsection{Datasets.}
We choose three real-world datasets, i.e. Diginetica, Gowalla, and Amazon home and kitchen, which (especially the first two) are commonly used in session-based recommendation, to evaluate the performance of different approaches. In particular, 
\emph{Diginetica} is from CIKM cup 2016 and consists of typical transactions data. Following \cite{li2017neural,liu2018stamp,qiu2019rethinking, wu2019session, ren2019repeatnet,chen2020handling}, we filter out items with less than $5$ interactions, and sessions with length smaller than $2$. Besides, we use around $80\%$ ad the train data, around $10$ as the validation data, and the sessions occurred in the last week as the test set (around $10\%$).
\emph{Gowalla} is a check-in dataset for point-of-interest (POI) recommendation. Following \cite{chen2020handling, guo2019streaming, tang2018personalized}, we keep the top $30,000$ most popular items, and treat a user's check-ins within a day as a session. Besides, we filter out sessions with length smaller than $2$ or larger than $20$. 
%
%
\emph{Amazon} contains product reviews and metadata from Amazon in home and kitchen category. We consider a user's interactions occurred in a day as a session, and filter out items with less than $5$ interactions and sessions with length smaller than $2$. 
For Gowalla and Amazon dataset, we sort sessions with increasing timestamp and take $70\%$ of sessions as the train set, $10\%$ as the validation set, the remaining $20\%$ as the test set. 
The statistics of the three datasets are summarized in Table \ref{tb:dataStatistics}.

\begin{table}[t]
\small\centering
\caption{Statistics of the three datasets.}\label{tb:dataStatistics}\vspace{-2mm}
\begin{tabular}{@{}c@{}ccc}
\toprule
Dataset& Diginetica& Gowalla &Amazon\\
\midrule
\#transactions & 982,961&1,122,788 &335,639\\
\#items & 43,097&29,510& 38,689\\
\#sessions& 780,328& 830,893& 246,661\\
average length &5.12&3.85& 3.77\\	
\#train sessions&647,370&592,481 &174,201\\	
\#validation sessions&72,100&83,080 &24,660\\
\#test sessions&60,858&155,332 &47,800\\
\bottomrule
\end{tabular}\vspace{-2mm}
\end{table}

\subsubsection{Baseline Models.}
We compare our CGSR with three traditional methods (\textbf{POP}, \textbf{ItemKNN} and \textbf{FPMC}), two RNN-based methods (\textbf{GRU4Rec}  and \textbf{NARM}), one attention-based method (\textbf{BERT4Rec}), and three state-of-the-art (SOTA) GNN-based methods (\textbf{SR-GNN}, \textbf{FGNN} and \textbf{LESSR}) for session-based recommendation.
\begin{itemize}
\item \textbf{POP} recommends the most popular items in the training set;
\item \textbf{ItemKNN} \cite{sarwar2001item} recommends items having the highest similarity with the last item of the session;
\item \textbf{FPMC} \cite{rendle2010factorizing} combines matrix factorization with the first-order MCs;
\item\textbf{GRU4Rec} \cite{hidasi2015session} stacks GRUs to process session data and tailors an ranking loss function to train the model;
\item \textbf{NARM} \cite{li2017neural} is a strong and solid RNN-based approach for SR, which utilizes vanilla attention to model the relationship of the last item with other items in a session to capture the main purpose;
\item \textbf{BERT4Rec} \cite{sun2019bert4rec} is a deep bidirectional sequential model with a Cloze objective loss;
\item \textbf{SR-GNN} \cite{wu2019session} uses a gated graph convolutional layer to obtain item embedding on session graph, and applies self-attention mechanism to last item embedding to obtain session representation; 
\item \textbf{FGNN} \cite{qiu2019rethinking} designs a novel model to collaboratively incorporate sequence order and latent order in the session graph;
\item \textbf{LESSR} \cite{chen2020handling} proposes a lossless encoding scheme and an edge-order preserving aggregation layer based on GRU, and designs a shortcut graph attention layer to effectively capture long-range dependencies among items.
\end{itemize}
Besides, we summarize the different variants of our CGSR model in Table \ref{tb:variants}.
\begin{table}[t]
\small\centering
 \caption{Descriptions of the  CGSR variants.}\label{tb:variants}\vspace{-2mm}
\begin{tabular}{l|l}
\toprule
CGSR variants  & Description\\
\hline
CGSR-ca  & remove both cause graph $\mathcal{G}^c$ and effect graph $\mathcal{G}^e$\\
CGSR-r & ignore correlation graph $\mathcal{G}^r$ \\
CGSR-p& not consider the session representation $\mathbf{S}^p$\\
CGSR\_mean &only consider $Score^p$ in Equation \ref{eq:Score}\\
CGSR-W  & edge weights in cause and effect graph are set to $1$\\
CGSR-CC & not rule out the impact of common cause\\
cause\_random & make the cause graph into a random graph\\
effect\_random & make the effect graph into a random graph\\
CGSR-chain & remove the ``chain'' relationship\\
CGSR-fork & ignore ``fork'' type\\
CGSR-collider &not consider second-order neighbors of collider type\\
CGSR-ca\_sec &consider the second-order relationship in cause and effect graphs\\
\bottomrule
\end{tabular}\vspace{-2mm}
\end{table}

\subsubsection{Evaluation Metrics.}
We adopt three widely used ranking-based metrics: \textbf{HR}@$K$, \textbf{MRR}@$K$ and \textbf{NDCG}@$K$\footnote{We only choose these three metrics because in next-item prediction, HR is identical to Recall, while MRR is identical to MAP (Mean Average Precision).}
to evaluate the recommendation accuracy, where a higher value indicates better performance, and $K$ is set to $5$, $10$ and $20$, respectively. \textbf{HR}@$K$ (Hit Ratio) denotes the hit ratio, i.e., the coverage rate of targeted predictions; \textbf{MRR}@$K$ (Mean Reciprocal Rank) indicates the ranking accuracy based on the ranking position of the recommended items (hits), and a larger value means the ground-truth items are ranked in the top of the ranked recommendation lists; \textbf{NDCG}@$K$ (Normalized Discounted Cumulative Gain) also rewards each hit based on its position in the ranked recommendation list.

\begin{table*}[htb]
\small\centering
\caption{Hyper-parameter setups of baselines.}\label{tb:baselineshyperParameter}\vspace{-2mm}
\setlength{\tabcolsep}{0.8mm}{
\begin{tabular}{c|l|l }
\hline
\textbf{Method} & \textbf{Datasets}   & \textbf{Hyper-parameter setups} \\
\hline
GRU4Rec  &Diginetica, Gowalla, Amazon	& GRU size=100, Batch size=32, Lr=0.2\\
\hline
NARM  &Diginetica, Gowalla, Amazon	&Embedding size=50, Batch size=512, Lr=0.001 \\
\hline
 \multirow{2}{*}{BERT4Rec}  &Diginetica, Amazon	&Embedding size=128, Batch size=512, Lr=0.001\\
 &Gowalla &Embedding size=64, Batch size=512, Lr=0.001\\
\hline
 \multirow{2}{*}{SR-GNN}  &Diginetica, Gowalla	&Embedding size=100, Batch size=100, Lr=0.001, $L_2$ penalty=1e-5\\
 &Amazon &Embedding size=170, Batch size=100, Lr=0.001, $L_2$ penalty=1e-5\\
 \hline
 \multirow{2}{*}{FGNN}  &Diginetica, Gowalla	&Embedding size=100, Batch size=100, Lr=0.001, $L_2$ penalty=1e-5 \\
 &Amazon	&Embedding size=150, Batch size=100, Lr=0.001, $L_2$ penalty=1e-5 \\
 \hline
 \multirow{3}{*}{LESSR}  &Diginetica	&Embedding size=32, Batch size=512, Lr=0.001, $L_2$ penalty=1e-4\\
 &Gowalla	&Embedding size=64, Batch size=512, Lr=0.001, $L_2$ penalty=1e-4\\
  &Amazon	&Embedding size=128, Batch size=512, Lr=0.001, $L_2$ penalty=1e-4\\
\hline
\end{tabular}}\\
\begin{tablenotes}
\item In the original papers, NARM, SR-GNN, FGNN, LESSR have used Diginetica; LESSR have processed Gowalla datasets.
Thus, for these scenarios, we directly implemented the corresponding settings.
\end{tablenotes}
\end{table*}

\subsubsection{Hyper-Parameter Setups}
Regarding each method, we empirically adopt the optimal hyper-parameter settings according to validation sessions on each dataset.
For the proposed CGSR,
we apply one layer WGAT in \emph{Item Representation Learner}, and use Adam optimizer with the initial learning rate (Lr) $0.001$ on Diginetica and Gowalla while $0.003$ on Amazon. The representation size of each item $d_0$ and $d$ are $110$ on Diginetica, $60$ on Gowalla, and $150$ on Amazon. All parameters are initialized using Gaussian distribution with a mean of $0$ and a standard deviation of $0.1$. The L2 penalty is set to $1e-6$ on Gowalla and $5e-6$ on Diginetica, Amazon. Moreover, the batch size is $20$ on Diginetica, $40$ on Gowalla, and $100$ on Amazon.
%
For baselines, we adopt the optimal settings mentioned in either the original papers for these datasets or the original codes.
The settings of baselines are shown in Table \ref{tb:baselineshyperParameter}.
Noted that for CGSR and the best baseline, for fair comparison, we run each experiment six times, report the average as the final result in Table \ref{tb:comparativeResults}, and conduct pair-wise t-test to validate the significance of the performance difference.

\subsection{Experimental Results}
Here, we present results to answer the first three RQs ($1-3$). 

\subsubsection{Effectiveness of CGSR over Baseline Methods (RQ1).}
\begin{table*}[htbp]
\setlength\tabcolsep{1pt}
\caption{Performance of all methods on three datasets in terms of $K=5,10,20$. The best performance is boldfaced, and the runner-up is underlined. We compute the improvements that CGSR achieves relative to the best baseline. Statistical significance of pairwise differences of CGSR vs. the best baseline is determined by a paired t-test ($^{*}$ for p-value $\leq$.05, $^{**}$ for p-value $\leq$.01, and $^{***}$ for p-value $\leq$.001).}\label{tb:comparativeResults}\vspace{-2mm}
\resizebox{\textwidth}{50mm}{
\begin{tabular}{c|l|ccc| cc|c|c cc c |l }
\hline
&&\multicolumn{3}{c}{Traditional}&\multicolumn{2}{|c|}{RNN-based}&\multicolumn{1}{|c|}{attention-based}&\multicolumn{4}{c|}{GNN-based}\\
\cline{3-12}
\textbf{Datasets} & \textbf{Metrics} & \textbf{POP} & \textbf{ItemKNN} & \textbf{FPMC} & \textbf{GRU4Rec} & \textbf{NARM}  & \textbf{BERT4Rec}& \textbf{SR-GNN} & \textbf{FGNN}  & \textbf{LESSR} & \textbf{CGSR}& \textbf{Improv.} \\
\hline
\multirow{9}{*}{Diginetica}
& HR@5 &0.0040	&0.1447	&0.1416	&0.1576	&0.2557  &0.2672     &0.2673  &0.2640  &\underline{0.2729} & \textbf{0.3230}& 18.36\%***	\\
&  HR@10  &0.0064	&0.2130	&0.1769	&0.2233	&0.3647 & 0.3823       &0.3774 &0.3792 &\underline{0.3852}  & \textbf{0.4399} &14.20\%***\\
 &  HR@20  &0.0970	&0.2897	&0.2573	&0.3004	&0.4889 &0.5087   &0.5076 &0.5113 &\underline{0.5147} &\textbf{0.5567} &8.16\%***\\
 & MRR@5  &0.0019	&0.0807	&0.0614	&0.0894	&0.1419  &0.1513    &0.1519 &0.1457 &\underline{0.1561} &\textbf{0.1823} &16.78\%***\\
&  MRR@10 &0.0022	&0.0897	&0.0663	&0.0981	&0.1581 &0.1665     &0.1665 &0.1610 &\underline{0.1708} &\textbf{0.1979} &15.87\%***\\
 &  MRR@20  &0.0024	&0.0965	&0.0707	&0.1034	&0.1647  &0.1752    &0.1755 &0.1702 &\underline{0.1799} &\textbf{0.2044} &13.62\%***\\
 &  NDCG@5  &0.0024	&0.0965	&0.0625	&0.1063	&0.1658  &0.1799   &0.1804 &0.1749   &\underline{0.1849} &\textbf{0.2172}&17.47\%***\\
 & NDCG@10  &0.0032	&0.1185	&0.0718	&0.1274	&0.2024 &0.2170    &0.2159 &0.2121 &\underline{0.2212} &\textbf{0.2549} &15.24\%***\\
 &  NDCG@20  &0.0040	&0.1379	&0.0788	&0.1469	&0.2315 &0.2489     &0.2488 &0.2455 &\underline{0.2538} &\textbf{0.2858}&12.61\%***\\

\hline
\multirow{9}{*}{Gowalla} 

& HR@5 &0.0183	&0.2614	&0.1869	&0.2874	&0.3506  &0.3547     	& 0.3557 &0.3471 &\underline{0.3577} & \textbf{0.3802} & 6.29\%***\\
&  HR@10   &0.0277	&0.3248	&0.2287	&0.3558	&0.4272 &0.4341    	& \underline{0.4359} &0.4281 & 0.4340 & \textbf{0.4612} & 5.80\%***\\
&  HR@20  &0.0500	&0.3891	&0.2834	&0.4326	&0.4989 &0.5127    	&\underline{0.5149} &0.5080 &0.5104 & \textbf{0.5389} & 4.66\%***\\
& MRR@5  &0.0090	&0.1718	&0.0976	&0.1863	&0.2209 &0.2381   	& 0.2383 &0.2212 & \underline{0.2403} & \textbf{0.2477} & 3.08\%*\\
&  MRR@10  &0.0102	&0.1803	&0.1089	&0.1954	&0.2312 &0.2487  	& 0.2490 &0.2321 &\underline{0.2505} & \textbf{0.2585} & 3.19\%***\\
&  MRR@20  &0.0118 &0.1847	&0.1116	&0.2006	&0.2345 &0.2541 	& 0.2546 &0.2376 & \underline{0.2557} & \textbf{0.2640} & 3.25\%***\\
&  NDCG@5  &0.0113	&0.1941	&0.1145	&0.2115	&0.2533 &0.2671 	& 0.2675 &0.2526 &\underline{ 0.2695} & \textbf{0.2808} & 4.19\%***\\
&  NDCG@10  &0.0143	&0.2146	&0.1239	&0.2336	&0.2782 &0.2928	& 0.2935 &0.2788 & \underline{0.2942} & \textbf{0.3070} &4.35\%***\\
 &  NDCG@20  &0.0200	&0.2309	&0.1348	&0.2530	&0.2977 &0.3127 	& \underline{0.3137} &0.2990 & 0.3135 & \textbf{0.3267} &4.14\%***\\

\hline
\multirow{9}{*}{Amazon} 
& HR@5 &0.0046	&0.0433	&0.0372	&0.0418	&0.0446  &0.0456  	& 0.0550 &0.0464 &\underline{0.0570} & \textbf{0.0607} &6.49\%***\\
&  HR@10   &0.0087	&0.0515	&0.0497	&0.0513	&0.0571 &0.0552 	& 0.0684 &0.0618 & \underline{0.0691} &\textbf{0.0709} & 2.60\%**\\
&  HR@20  &0.0160	&0.0591	&0.0589	&0.0649	&0.0718 &0.0661 	&\underline{0.0816} &0.0781 &0.0813 & \textbf{0.0835} & 2.33\%*\\
& MRR@5  &0.0021	&0.0288	&0.0266	&0.0283	&0.0300 &0.0329 	& 0.0377 &0.0304 & \underline{0.0396} &\textbf{0.0440} & 11.11\%***\\
&  MRR@10  &0.0026	&0.0299	&0.0281	&0.0297	&0.0317  &0.0341  	& 0.0394 &0.0325 & \underline{0.0412} &\textbf{0.0457} & 10.92\%***\\
&  MRR@20  &0.0031 &0.0304	&0.0288	&0.0303	&0.0327 &0.0349 	& 0.0403 &0.0336 & \underline{0.0420} &\textbf{0.0466} & 10.95\%***\\
&  NDCG@5  &0.0027	&0.0325	&0.0295	&0.0312	&0.0336 &0.0361 	& 0.0420 &0.0344 & \underline{ 0.0439} &\textbf{0.0478} & 8.88\%***\\
&  NDCG@10  &0.0040	&0.0351	&0.0313	&0.0339	&0.0377 &0.0391 	& 0.0463 &0.0394 & \underline{0.0478} &\textbf{0.0519} &8.58\%***\\
 &  NDCG@20  &0.0059	&0.0370	&0.0346	&0.0369	&0.0414 &0.0419 	& 0.0496 &0.0435 & \underline{0.0509} &\textbf{0.0546} &7.27\%***\\
\hline
\end{tabular}}\vspace{-2mm}
\end{table*}

To demonstrate the overall performance of CGSR, we compare it with SOTA baseline methods. The comparative results on the three datasets are present in Table \ref{tb:comparativeResults}, where we have some interesting observations as below: 
(1) The DL-based methods (both RNN-based and GNN-based) generally perform better than traditional methods, demonstrating the capability of DL techniques on processing session data for effective recommendation. Particularly, among the traditional methods, ItemKNN, which is grounded on the similarity between items within a session, performs much better than POP, and slightly better than first-order MC-based FPMC; (2) Across the DL-based methods, GNN-based methods generally outperform RNN-based methods, which validates the effectiveness of GNN models and graph data structures for SR. For the two RNN-based methods, NARM performs better than GRU4Rec, and its performance is comparable to other GNN-based methods; and (3) The performance of CGSR is significantly better than SOTA GNN-based methods, validating its effectiveness of distinguishing causality relationship between items from correlation relationship. Among all the GNN-based baselines, LESSR performs the best as it captures both local and long-range dependencies among items. 

From Table \ref{tb:comparativeResults}, we also observe that the improvements of CGSR on Diginetica and Amazon are both larger than those on Gowalla. This can be partially explained by Figures \ref{fig:digineticaStatistics} and \ref{fig:causalityStatistics}  which show that stronger causality relationship between items is more prevalent on Diginetica (Amazon) than that on Gowalla. Besides, Gowalla relates to location-based social networking where users share their locations by check-ins. In this case, the directed relationship between check-ins might not so significant compared to that on typical transaction datasets like Diginetica and Amazon.

\begin{table}[htb]
\small\centering
 \caption{Impact of causality and correlation.}\label{tb:causalityEffectiveness}\vspace{-2mm}
\setlength{\tabcolsep}{0.8mm}{
\begin{tabular}{c|c|ccccc }
\toprule
\textbf{Datasets} & \textbf{Metrics}   & \textbf{CGSR-ca} & \textbf{CGSR-r} & \textbf{CGSR-p} & \textbf{CGSR\_mean}  & \textbf{CGSR} \\
\hline
\multirow{3}{*}{Diginetica}
& HR@5      &0.2950	&0.2663	&0.3171	&0.2641  &\textbf{0.3230}  \\
&  HR@10    &0.4054	&0.3811	&0.4349	&0.3787  &\textbf{0.4399}  \\
&  HR@20    &0.5173	&0.5117	&0.5544	&0.5103  &\textbf{0.5567}  \\
& MRR@5     &0.1644	&0.1408	&0.1771	&0.1386  &\textbf{0.1823}  \\
&  MRR@10   &0.1784	&0.1561	&0.1940	&0.1507  &\textbf{0.1979}  \\
&  MRR@20   &0.1873	&0.1687	&0.2002	&0.1651  &\textbf{0.2044}  \\
&  NDCG@5   &0.1952	&0.1744	&0.2108	&0.1712  &\textbf{0.2172}  \\
&  NDCG@10  &0.2321	&0.2115	&0.2492	&0.2105  &\textbf{0.2549}  \\
 &  NDCG@20 &0.2612	&0.2427	&0.2794	&0.2408  &\textbf{0.2858}  \\


\hline
\multirow{3}{*}{Gowalla}
& HR@5  &0.3718	&0.3601 &0.3798      &0.3561   & \textbf{0.3802}\\
&  HR@10    &0.4516	&0.4388 &0.4601  &0.4332  & \textbf{0.4612}\\
&  HR@20    &0.5272	&0.5183 &0.5386  &0.5041  & \textbf{0.5389}\\
 & MRR@5     &0.2399	&0.2338 &0.2467 &0.2304  & \textbf{0.2477}\\
 &  MRR@10   &0.2451	&0.2444 &0.2575 &0.2397  & \textbf{0.2585}\\
 &  MRR@20   &0.2532	&0.2499 &0.2629 &0.2456  & \textbf{0.2640}\\
&  NDCG@5   &0.2708	&0.2653 &0.2802  &0.2614  & \textbf{0.2808}\\
 &  NDCG@10  &0.2954	&0.2908 &0.3063 &0.2873  & \textbf{0.3070}\\
 &  NDCG@20  &0.3143	&0.3109 &0.3258 &0.3054  & \textbf{0.3267}\\
 
 \hline
\multirow{3}{*}{Amazon}
& HR@5  &0.0560	&0.0523 &0.0547         &0.0515  & \textbf{0.0607}\\ 
&  HR@10    &0.0687	    &0.0641 &0.0675 &0.0627  & \textbf{0.0709}\\ 
&  HR@20    &0.0815	&0.0773 &0.0801     &0.0758  & \textbf{0.0835}\\  
 & MRR@5     &0.0373	&0.0365 &0.0390    &0.0352  & \textbf{0.0440}\\  
 &  MRR@10   &0.0408	&0.0389 &0.0416    &0.0371  & \textbf{0.0457}\\  
  &  MRR@20   &0.0433	&0.0416 &0.0434   &0.0410  & \textbf{0.0466}\\  
&  NDCG@5   &0.0427	    &0.0401 &0.0429 &0.0390  & \textbf{0.0478}\\ 
 &  NDCG@10  &0.0475	&0.0446 &0.0473    &0.0433  & \textbf{0.0519}\\   
&  NDCG@20  &0.0505	&0.0473 &0.0504    &0.0460  & \textbf{0.0546}\\   
\bottomrule
\end{tabular}}\vspace{-2mm}
\end{table}

\subsubsection{Effectiveness of Causality Graphs vs. Correlation Graph (RQ2).}
CGSR considers both causality and correlation graphs. To explore the effectiveness of each type of relationship, we compare CGSR with four variants: (1) CGSR-ca removes both cause graph $\mathcal{G}^c$ and effect graph $\mathcal{G}^e$; (2) CGSR-r ignores correlation graph $\mathcal{G}^r$; (3) CGSR-p does not consider the session representation $\mathbf{S}^p$; and (4) CGSR\_mean only considers $Score^p$ in Equation \ref{eq:Score}. The performance of CGSR and the four variants are shown in Table \ref{tb:causalityEffectiveness}. 

As shown in Table \ref{tb:causalityEffectiveness}, CGSR performs superior to the four variants across all metrics, validating the effectiveness of the four designs, particularly distinguishing directed causality relationship from undirected correlation relationship between items.
Besides, CGSR-ca performs better than CGSR-r, implying that only considering correlation graph is better than only considering causality relationship. This might be due to the undirected correlation graph might already cover the causality relationship but vice versa not.

\begin{table}[htb]
\small\centering
\caption{Performance of different causality weights.}\label{tb:causalityWeights}\vspace{-2mm}
\setlength{\tabcolsep}{0.8mm}{
\begin{tabular}{c|c|ccc }
\toprule
\textbf{Datasets} & \textbf{Metrics}   & \textbf{CGSR-W} & \textbf{CGSR-CC} & \textbf{CGSR} \\
\hline
\multirow{3}{*}{Diginetica}

& HR@5   &0.3192	&0.3168	&\textbf{0.3230}\\
& HR@10  &0.4351	&0.4332	&\textbf{0.4399}\\
&  HR@20 &0.5515	&0.4982	&\textbf{0.5567}\\
& MRR@5  &0.1773	&0.1782	&\textbf{0.1823}\\
&  MRR@10 &0.1917	&0.1925	&\textbf{0.1979}\\
&  MRR@20 &0.2003	&0.2007	&\textbf{0.2044}\\
&  NDCG@5 &0.2101	&0.2113	&\textbf{0.2172}\\
&  NDCG@10 &0.2506	&0.2524	&\textbf{0.2549}\\
 &  NDCG@20 &0.2786	&0.2794	&\textbf{0.2858}\\


\hline
\multirow{3}{*}{Gowalla}
& HR@5    &0.3797 &0.3795 & \textbf{0.3802}\\
&  HR@10   &0.4601 &0.4607 & \textbf{0.4612}\\
&  HR@20   &0.5375 &0.5379 & \textbf{0.5389}\\
& MRR@5    &0.2471 &0.2470 & \textbf{0.2477}\\
&  MRR@10  &0.2581 &0.2580 & \textbf{0.2585}\\
&  MRR@20  &0.2635 &0.2634 & \textbf{0.2640}\\
&  NDCG@5  &0.2807 &0.2801 & \textbf{0.2808}\\
&  NDCG@10 &0.3068 &0.3064 & \textbf{0.3070}\\
 &  NDCG@20 &0.3264 &0.3263 & \textbf{0.3267}\\

\bottomrule
\end{tabular}}\vspace{-2mm}
\end{table}

\begin{figure}[htb]
    \centering
	\footnotesize
	\begin{tikzpicture}
      \matrix[
          matrix of nodes,
          draw,
          inner sep=0.1em,
          ampersand replacement=\&,
          font=\scriptsize,
          anchor=south
        ]
        { 
        \ref{plots:tplot1} cause\_random
		\ref{plots:tplot2} effect\_random
		\ref{plots:tplot3} CGSR\\
          };
    \end{tikzpicture}\\
	
	\begin{tikzpicture}
	\begin{groupplot}[group style={
		group name=myplot,
		group size= 3 by 3,  horizontal sep=1.2cm}, 
		height=4cm, width=4cm,
	ylabel style={yshift=-0.1cm},
	every tick label/.append style={font=\small}
	]

	\nextgroupplot[ybar=0.10,
	bar width=0.55em,
	ylabel={HR@5},
	scaled ticks=false,
	yticklabel style={/pgf/number format/.cd,fixed,precision=3},
	ymin=0.2, ymax=0.41,
	enlarge x limits=0.4,
	symbolic x coords={Diginetica,Gowalla},
	ylabel style = {font=\small},
	xtick=data,
	ytick={0.20,0.30,0.40},
	]
	\addplot coordinates {
		(Diginetica,0.2691) (Gowalla, 0.3254)};\label{plots:tplot1}
	\addplot coordinates {
		(Diginetica,0.2725) (Gowalla, 0.3272)};\label{plots:tplot2}
	\addplot coordinates {
		(Diginetica,0.3230) (Gowalla, 0.3802)};\label{plots:tplot3}

	\nextgroupplot[ybar=0.10,
	bar width=0.55em,
	ylabel={MRR@5},
	scaled ticks=false,
	yticklabel style={/pgf/number format/.cd,fixed,precision=3},
	ymin=0.1, ymax=0.31,
	enlarge x limits=0.4,
	symbolic x coords={Diginetica,Gowalla},
	ylabel style = {font=\small},
	xtick=data,
	ytick={0.1,0.20,0.3},
	]
	\addplot coordinates {
		(Diginetica,0.1501) (Gowalla, 0.1981)};\label{plots:tplot1}
	\addplot coordinates {
		(Diginetica,0.1512) (Gowalla, 0.2006)};\label{plots:tplot2}
	\addplot coordinates {
		(Diginetica,0.1823) (Gowalla, 0.2477)};\label{plots:tplot3}
		
	\nextgroupplot[ybar=0.10,
	bar width=0.55em,
	ylabel={NDCG@5},
	scaled ticks=false,
	yticklabel style={/pgf/number format/.cd,fixed,precision=3},
	ymin=0.1, ymax=0.31,
	enlarge x limits=0.4,
	symbolic x coords={Diginetica,Gowalla},
	ylabel style = {font=\small},
	xtick=data,
	ytick={0.1,0.2,0.3},
	]
	\addplot coordinates {
		(Diginetica,0.1653) (Gowalla, 0.2359)};\label{plots:tplot1}
	\addplot coordinates {
		(Diginetica,0.1657) (Gowalla, 0.2362)};\label{plots:tplot2}
	\addplot coordinates {
		(Diginetica,0.2172) (Gowalla, 0.2808)};\label{plots:tplot3}

	\nextgroupplot[ybar=0.10,
	bar width=0.55em,
	ylabel={HR@10},
	scaled ticks=false,
	yticklabel style={/pgf/number format/.cd,fixed,precision=3},
	ymin=0.3, ymax=0.51,
	enlarge x limits=0.4,
	symbolic x coords={Diginetica,Gowalla},
	ylabel style = {font=\small},
	xtick=data,
	ytick={0.3,0.40,0.5},
	]
	\addplot coordinates {
		(Diginetica,0.3560) (Gowalla, 0.3959)};\label{plots:tplot1}
	\addplot coordinates {
		(Diginetica,0.3565) (Gowalla, 0.3978)};\label{plots:tplot2}
	\addplot coordinates {
		(Diginetica,0.4399) (Gowalla, 0.4612)};\label{plots:tplot3}

	\nextgroupplot[ybar=0.10,
	bar width=0.55em,
	ylabel={MRR@10},
	scaled ticks=false,
	yticklabel style={/pgf/number format/.cd,fixed,precision=3},
	ymin=0.1, ymax=0.31,
	enlarge x limits=0.4,
	symbolic x coords={Diginetica,Gowalla},
	ylabel style = {font=\small},
	xtick=data,
	ytick={0.1,0.2,0.3},
	]
	\addplot coordinates {
		(Diginetica,0.1519) (Gowalla, 0.2114)};\label{plots:tplot1}
	\addplot coordinates {
		(Diginetica,0.1531) (Gowalla, 0.2118)};\label{plots:tplot2}
	\addplot coordinates {
		(Diginetica,0.1979) (Gowalla, 0.2585)};\label{plots:tplot3}
		
	\nextgroupplot[ybar=0.10,
	bar width=0.55em,
	ylabel={NDCG@10},
	scaled ticks=false,
	yticklabel style={/pgf/number format/.cd,fixed,precision=3},
	ymin=0.15, ymax=0.36,
	enlarge x limits=0.4,
	symbolic x coords={Diginetica,Gowalla},
	ylabel style = {font=\small},
	xtick=data,
	ytick={0.15,0.25,0.35},
	]
	\addplot coordinates {
		(Diginetica,0.1997) (Gowalla, 0.2555)};\label{plots:tplot1}
	\addplot coordinates {
		(Diginetica,0.2007) (Gowalla, 0.2563)};\label{plots:tplot2}
	\addplot coordinates {
		(Diginetica,0.2549) (Gowalla, 0.3070)};\label{plots:tplot3}

	\nextgroupplot[ybar=0.10,
	bar width=0.55em,
	ylabel={HR@20},
	scaled ticks=false,
	yticklabel style={/pgf/number format/.cd,fixed,precision=3},
	ymin=0.4, ymax=0.61,
	enlarge x limits=0.4,
	symbolic x coords={Diginetica,Gowalla},
	ylabel style = {font=\small},
	xtick=data,
	ytick={0.4,0.5,0.6},
	]
	\addplot coordinates {
		(Diginetica,0.4788) (Gowalla, 0.4707)};\label{plots:tplot1}
	\addplot coordinates {
		(Diginetica,0.4796) (Gowalla, 0.4720)};\label{plots:tplot2}
	\addplot coordinates {
		(Diginetica,0.5567) (Gowalla, 0.5389)};\label{plots:tplot3}

    \nextgroupplot[ybar=0.10,
	bar width=0.55em,
	ylabel={MRR@20},
	scaled ticks=false,
	yticklabel style={/pgf/number format/.cd,fixed,precision=3},
	ymin=0.1, ymax=0.32,
	enlarge x limits=0.4,
	symbolic x coords={Diginetica,Gowalla},
	xtick=data,
	ytick={0.1,0.2,0.3},
	]
	\addplot coordinates {
		(Diginetica,0.1604) (Gowalla, 0.2165)};
	\addplot coordinates {
		(Diginetica,0.1617) (Gowalla, 0.2170)};
	\addplot coordinates {
		(Diginetica,0.2044) (Gowalla, 0.2640)};

    \nextgroupplot[ybar=0.10,
	bar width=0.55em,
	ylabel={NDCG@20},
	scaled ticks=false,
	yticklabel style={/pgf/number format/.cd,fixed,precision=3},
	ymin=0.2, ymax=0.42,
	enlarge x limits=0.4,
	symbolic x coords={Diginetica,Gowalla},
	xtick=data,
	ytick={0.2,0.3,0.4},
	]
	\addplot coordinates {
		(Diginetica,0.2307) (Gowalla, 0.2744)};
	\addplot coordinates {
		(Diginetica,0.2318) (Gowalla, 0.2751)};
	\addplot coordinates {
		(Diginetica,0.2858) (Gowalla, 0.3267)};
	\end{groupplot}
    \end{tikzpicture}\vspace{-2mm}
    \caption{Impact of casue or effect graph.}\vspace{-2mm}
    \label{fig:random}
\end{figure}
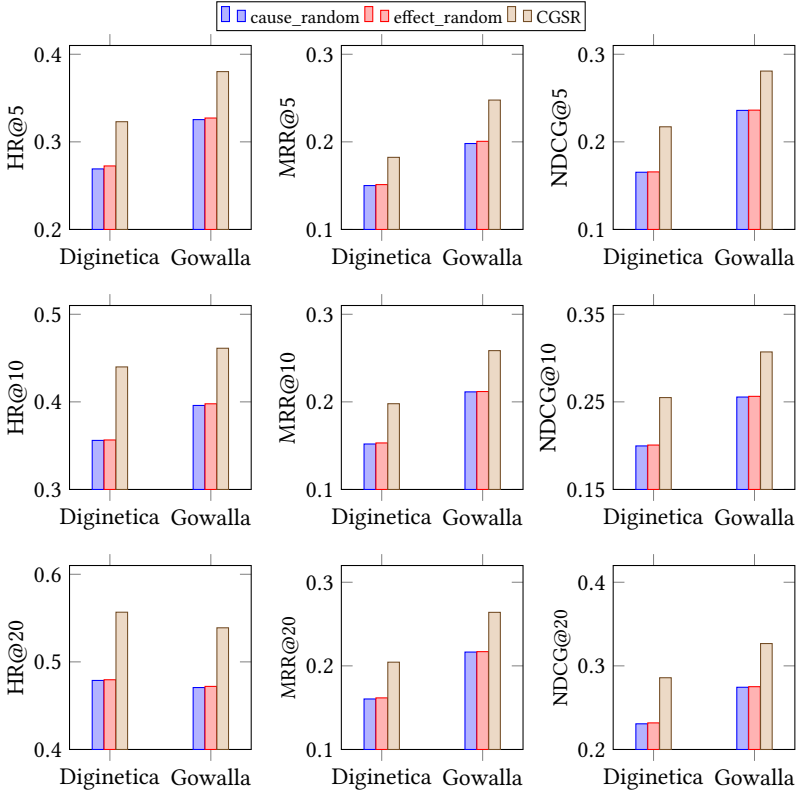

Furthermore, we also explore the effectiveness of weights designs in cause and effect graphs (i.e., \textbf{impact of causality weights}). CGSR defines weights of causality-related graphs as discussed in Section \ref{sec:prelim}. To validate the effectiveness, we compare CGSR with two alternatives: (1) CGSR-W, where edge weights in cause and effect graph are set to $1$; and (2) CGSR-CC, which does not rule out the impact of common cause. As can 
see in Table \ref{tb:causalityWeights}, CGSR performs better than the two variants, verifying the effectiveness of our design and the necessity of removing common cause in identifying causality relationship.

To further validate the impact of cause or effect graphs, we make the cause graph and effect graph into a random graph (cause\_random and effect\_random), respectively. The results are shown in Figure \ref{fig:random}, where we can see that cause\_random and effect\_random perform consistently worse than CGSR, also implying the effectiveness of building cause and effect graphs.

\subsubsection{Ablation Study (RQ2).} Besides causality graphs, we have some other innovative designs in CGSR: (1) three types of second-order relationship in building correlation graph; (2) first-order relationship in building cause and effect graphs; and (3) adopting one layer WGAT in Item Representation Learner.

Towards the first issue, to explore \textbf{the effectiveness of the second-order relationship}, we compare our model with three variants: (1) CGSR-chain removes the ``chain'' relationship; (2) CGSR-fork ignores ``fork'' type; and (3) CGSR-collider does not consider second-order neighbors of collider type. Table \ref{tb:second-orderRelationship} summarizes the comparative results and shows that all the three types contribute to performance improvement, but fork relation is less significant than the other two. This is consistent with directed graphical model principles which indicate that $v_i$ and $v_j$ in Figure \ref{fig:secondRelationships} incline to be independent with each other given a known $v_k$ in fork pattern \cite{koller2009probabilistic}.

\begin{table}[htb]
\small\centering
 \caption{Impact of second-order relationship in correlation graph.}\label{tb:second-orderRelationship}\vspace{-2mm}
\setlength{\tabcolsep}{0.8mm}{
\begin{tabular}{c|c|cccc }
\toprule
\textbf{Datasets} & \textbf{Metrics}   & \textbf{CGSR-chain} & \textbf{CGSR-fork} & \textbf{CGSR-collider} & \textbf{CGSR} \\
\hline
\multirow{3}{*}{Diginetica}

& HR@5 &0.3149	&0.3153	&0.3129	& \textbf{0.3230}\\
&  HR@10 &0.4323	&0.4339	&0.4318	& \textbf{0.4399}\\
&  HR@20  &0.5507	&0.5519	&0.5497	& \textbf{0.5567}\\
& MRR@5 &0.1783	&0.1789	&0.1775	& \textbf{0.1823}\\
&  MRR@10 &0.1900	&0.1907	&0.1894	& \textbf{0.1979}\\
&  MRR@20 &0.1984	&0.1992	&0.1978	& \textbf{0.2044}\\
&  NDCG@5&0.2092	&0.2097	&0.2081	& \textbf{0.2172}\\
&  NDCG@10&0.2472	&0.2480	&0.2465	& \textbf{0.2549}\\
 &  NDCG@20 &0.2777	&0.2788	&0.2769	& \textbf{0.2858}\\


\hline
\multirow{3}{*}{Gowalla}
& HR@5  &0.3754   &0.3780  &0.3759  & \textbf{0.3802}\\
&  HR@10    &0.4552   &0.4596  &0.4560  & \textbf{0.4612}\\
&  HR@20     &0.5327   &0.5372  &0.5335  & \textbf{0.5389}\\
 & MRR@5      &0.2457   &0.2459  &0.2450  & \textbf{0.2477}\\
 &  MRR@10    &0.2564   &0.2569  &0.2557  & \textbf{0.2585}\\
 &  MRR@20    &0.2617   &0.2623  &0.2611  & \textbf{0.2640}\\
&  NDCG@5    &0.2781   &0.2788  &0.2776  & \textbf{0.2808}\\
 &  NDCG@10   &0.3039   &0.3053  &0.3036  & \textbf{0.3070}\\
 &  NDCG@20   &0.3235   &0.3249  & 0.3232  & \textbf{0.3267}\\

\bottomrule
\end{tabular}}\vspace{-2mm}
\end{table}

Towards the second issue, we compare CGSR with the variant: CGSR-ca\_sec, which considers the second-order relationship in cause and effect graphs. The results of CGSR and the variant are shown in Table \ref{tb:ca_second}. From Table \ref{tb:ca_second}, CGSR performs superior to the variant across all metrics, partially validating the effectiveness of only considering first-order relationship in our study.

\begin{table}[htb]
\small\centering
  \caption{Impact of first-order relationship in cause and effect graphs.}\label{tb:ca_second}\vspace{-2mm}
\setlength{\tabcolsep}{0.8mm}{
\begin{tabular}{c|c|cc }
\toprule
\textbf{Datasets} & \textbf{Metrics}   & \textbf{CGSR-ca\_sec}  & \textbf{CGSR} \\
\hline
\multirow{3}{*}{Diginetica}

& HR@5   &0.3132	& \textbf{0.3230}\\
&  HR@10 &0.4320	& \textbf{0.4399}\\
&  HR@20 &0.5504	& \textbf{0.5567}\\
& MRR@5  &0.1778	& \textbf{0.1823}\\
&  MRR@10 &0.1897	& \textbf{0.1979}\\
&  MRR@20 &0.1979	& \textbf{0.2044}\\
&  NDCG@5 &0.2085	& \textbf{0.2172}\\
&  NDCG@10 &0.2469	& \textbf{0.2549}\\
 &  NDCG@20 &0.2770	& \textbf{0.2858}\\


\hline
\multirow{3}{*}{Gowalla}
& HR@5  &0.3624 & \textbf{0.3802}\\
&  HR@10    &0.4448  & \textbf{0.4612}\\
&  HR@20     &0.5254  & \textbf{0.5389}\\
 & MRR@5      &0.2273  & \textbf{0.2477}\\
 &  MRR@10    &0.2384  & \textbf{0.2585}\\
 &  MRR@20    &0.2440  & \textbf{0.2640}\\
&  NDCG@5    &0.2610 & \textbf{0.2808}\\
 &  NDCG@10   &0.2878  & \textbf{0.3070}\\
 &  NDCG@20   &0.3081 & \textbf{0.3267}\\

\bottomrule
\end{tabular}}\vspace{-2mm}
\end{table}

Towards the third issue, other models like gated graph neural network (GGNN) \cite{li2015gated} are also suitable for directed and weighted graphs. To verify the validity of one layer WGAT for CGSR, we consider model variants by varying the number of GGNN layers and WGAT layers, respectively. 
The comparative results regarding these variants are summarized in Table \ref{tb:differentGNN}, which shows that WGAT-related variants performs better than GGNN-related variants, and one layer WGAT consistently performs better across all scenarios, demonstrating the effectiveness of our design in \emph{Item Representation Learner}.

\begin{table}[htb]
\small\centering
\caption{Performance with different GNN layers.}\label{tb:differentGNN}\vspace{-2mm}
\setlength{\tabcolsep}{0.8mm}{
\begin{tabular}{c|c|ccccccc }
\toprule
\textbf{Datasets} & \textbf{Metrics}   & \textbf{3*GGNN} & \textbf{2*GGNN} & \textbf{GGNN} & \textbf{3*WGAT} & \textbf{2*WGAT} & \textbf{WGAT+GGNN} & \textbf{WGAT}\\
\hline
\multirow{3}{*}{Diginetica}

& HR@5 & 0.2751	& 0.2762	&0.2761	&0.2807	&0.2913	&0.3064	& \textbf{0.3230}\\
&  HR@10  & 0.3907	& 0.3918	&0.3914	&0.4011	&0.4123	&0.4241	& \textbf{0.4399}\\
&  HR@20 & 0.5153	& 0.5161	&0.5155	&0.5279	&0.5389	&0.5439	& \textbf{0.5567}\\
& MRR@5 & 0.1576	& 0.1588	&0.1580	&0.1566	&0.1587	&0.1687	& \textbf{0.1823}\\
&  MRR@10 & 0.1706	& 0.1721	&0.1720	&0.1728	&0.1756	&0.1846	& \textbf{0.1979}\\
&  MRR@20 & 0.1806	& 0.1819	&0.1809	&0.1813	&0.1831	&0.1912	& \textbf{0.2044}\\
&  NDCG@5& 0.1857	& 0.1863	&0.1857	&0.1866	&0.1908	&0.2017	& \textbf{0.2172}\\
&  NDCG@10& 0.2248	& 0.2250	&0.2247	&0.2257	&0.2307	&0.2406	& \textbf{0.2549}\\
 &  NDCG@20& 0.2550	& 0.2551	&0.2548	&0.2571	&0.2615	&0.2710	& \textbf{0.2858}\\


\hline
\multirow{3}{*}{Gowalla}
& HR@5   &0.3570  &0.3652  &0.3471   &0.3489  &0.3613  &0.3730 & \textbf{0.3802}\\
&  HR@10     &0.4370  &0.4438  &0.4270   &0.4276  &0.4425  &0.4547 & \textbf{0.4612}\\
&  HR@20      &0.5167  &0.5219  &0.5081   &0.5056  &0.5213  &0.5349 & \textbf{0.5389}\\
 & MRR@5       &0.2322  &0.2392  &0.2270 &0.2209  &0.2311&0.2432  & \textbf{0.2477}\\
 &  MRR@10     &0.2429  &0.2497  &0.2377   &0.2315  &0.2420  &0.2541 & \textbf{0.2585}\\
 &  MRR@20    &0.2484  &0.2551  &0.2434  &0.2369  &0.2474  &0.2597 & \textbf{0.2640}\\
&  NDCG@5     &0.2633  &0.2706  &0.2569   &0.2529  &0.2635  &0.2756 & \textbf{0.2808}\\
 &  NDCG@10   &0.2892  &0.2860  &0.2828 &0.2783  &0.2899  &0.3020 & \textbf{0.3070}\\
 &  NDCG@20   &0.3093  &0.3158  &0.3043   &0.2981  &0.3098  &0.3223 & \textbf{0.3267}\\

\bottomrule
\end{tabular}}\vspace{-2mm}
\end{table}

\subsubsection{Sensitivity of Hyper-parameters (RQ3).}
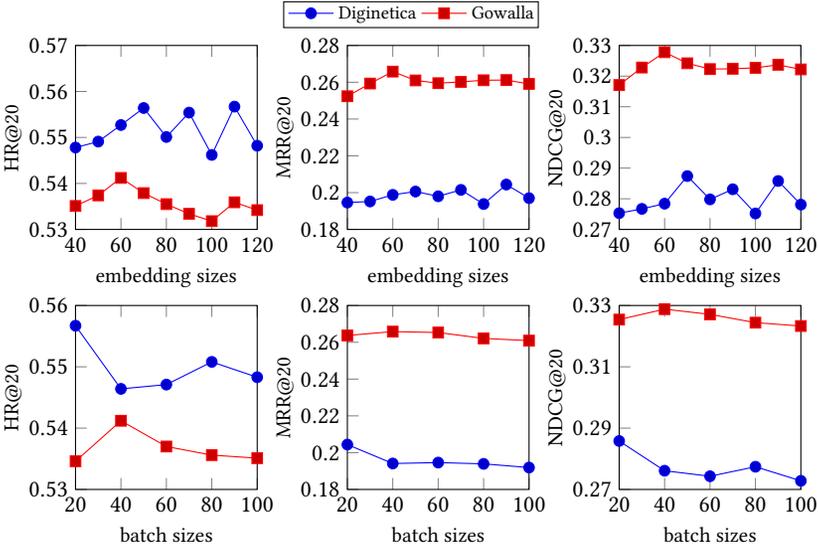
\begin{figure*}[htbp]
    \centering
	\footnotesize
	\begin{tikzpicture}
      \matrix[
          matrix of nodes,
          draw,
          inner sep=0.1em,
          ampersand replacement=\&,
          font=\scriptsize,
          anchor=south
        ]
        { 
		\ref{plots:Dsize1} Diginetica
		\ref{plots:Gsize2} Gowalla\\
          };
    \end{tikzpicture}\\
\begin{tikzpicture}
	\begin{groupplot}[group style={
		group name=myplot,
		group size= 3 by 2,  horizontal sep=1.2cm}, 
		height=4cm, width=4cm,
	ylabel style={yshift=-0.1cm},
	every tick label/.append style={font=\small}
	]
	
\nextgroupplot[
ylabel=HR@20,
xlabel=embedding sizes,
xmin=40,
xmax=120,
ymin=0.53,
ymax=0.57,
ytick={0.53,0.54, 0.55, 0.56, 0.57},
ytick pos=left
]
\addplot coordinates {
(40,0.5478) 
(50,0.5491) 
(60, 0.5527)
(70, 0.5564)
(80, 0.5501)
(90, 0.5554)
(100, 0.5462)
(110,0.5567)
(120,0.5482)
(130,0.5608)
};\label{plots:Dsize1}
\addplot coordinates {
(40,0.5351)
(50,0.5374)
(60,0.5412)
(70,0.5379)
(80,0.5355)
(90,0.5334)
(100,0.5318)
(110,0.5359)
(120,0.5342)
(130,0.5336)
};\label{plots:Gsize2}

\nextgroupplot[
ylabel=MRR@20,
xlabel=embedding sizes,
xmin=40,
xmax=120,
ymin=0.18,
ymax=0.28,
ytick={0.18,0.20, 0.22, 0.24, 0.26, 0.28},
ytick pos=left
]
\addplot coordinates {
(40,0.1946)
(50,0.1952)
(60,0.1988)
(70,0.2006)
(80,0.1980)
(90,0.2015)
(100,0.1938)
(110,0.2044)
(120,0.1970)
(130,0.2025)
};
\addplot coordinates {
(40,0.2524)
(50,0.2593)
(60,0.2658)
(70,0.2610)
(80,0.2595)
(90,0.2602)
(100,0.2611)
(110,0.2612)
(120,0.2591)
(130,0.2570)
};

\nextgroupplot[
ylabel=NDCG@20,
xlabel=embedding sizes,
xmin=40,
xmax=120,
ymin=0.27,
ymax=0.33,
ytick={0.27,0.28, 0.29, 0.30, 0.31, 0.32, 0.33},
ytick pos=left
]
\addplot coordinates {
(40,0.2753)
(50, 0.2767)
(60, 0.2784)
(70, 0.2874)
(80, 0.2798)
(90,0.2831)
(100,0.2752)
(110, 0.2858)
(120,0.2781)
(130, 0.2835)
};
\addplot coordinates {
(40,0.3171)
(50,0.3228)
(60,0.3278)
(70,0.3242)
(80,0.3223)
(90,0.3224)
(100,0.3227)
(110,0.3237)
(120,0.3222)
(130,0.3219)
};

\nextgroupplot[
ylabel=HR@20,
xlabel=batch sizes,
xmin=20,
xmax=100,
ymin=0.53,
ymax=0.56,
ytick={0.53,0.54, 0.55, 0.56},
ytick pos=left
]
\addplot coordinates {
(20,0.5567) 
(40,0.5464) 
(60, 0.5471) 
(80,0.5508) 
(100,0.5483) 

};
\addplot coordinates {
(20,0.5346) 
(40,0.5412)
(60,0.5370)
(80,0.5356)
(100,0.5351)
};

\nextgroupplot[
ylabel=MRR@20,
xlabel=batch sizes,
xmin=20,
xmax=100,
ymin=0.18,
ymax=0.28,
ytick={0.18,0.20, 0.22, 0.24, 0.26, 0.28},
ytick pos=left
]
\addplot coordinates {
(20,0.2044)
(40,0.1941)
(60,0.1946)
(80,0.1939)
(100,0.1919)  
};
\addplot coordinates {
(20,0.2637)
(40,0.2658)
(60,0.2653)
(80,0.2621)
(100,0.2609)
};

\nextgroupplot[
ylabel=NDCG@20,
xlabel=batch sizes,
xmin=20,
xmax=100,
ymin=0.27,
ymax=0.33,
ytick={0.27, 0.29, 0.31, 0.33},
ytick pos=left
]
\addplot coordinates {
(20, 0.2858)
(40, 0.2761)
(60, 0.2743)
(80, 0.2774)
(100,0.2728)
};
\addplot coordinates {
(20,0.3254)
(40,0.3288)
(60,0.3271)
(80,0.3244)
(100,0.3233)
};

\end{groupplot}

\end{tikzpicture}\vspace{-2mm}
   \caption{Model performance of different embedding sizes and batch sizes ($K=20$).}\vspace{-2mm}
    \
\label{fig:embeddingBatchSize}
\end{figure*}

We investigate the impact of embedding size $d$ and batch size on CGSR model, by deploying a grid search in the range of $\{40, 50, 60, 70, 80, 90, 100, 110, 120\}$ and $\{20, 40, 60, 80, 100\}$ for $d$ and batch size, respectively. Figure \ref{fig:embeddingBatchSize} shows the experiment results. Generally speaking, our method is comparatively insensitive to the two hyper-parameters. Besides, while varying the hyper-parameters, the performance on Diginetica fluctuates more obviously, which is consistent with our previous analysis that causality patterns are more prevalent on Diginetica than on Gowalla.

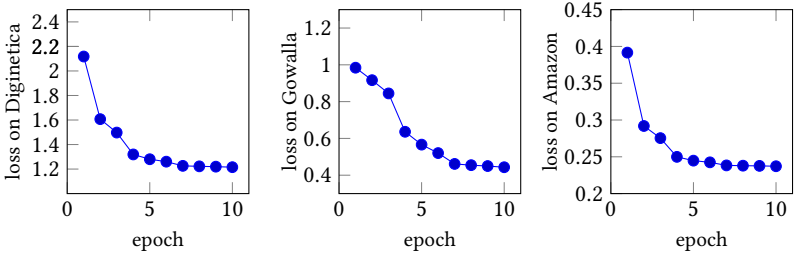
\begin{figure*}[htbp]
    \centering
	\footnotesize

\begin{tikzpicture}
	\begin{groupplot}[group style={
		group name=myplot,
		group size= 3 by 1,  horizontal sep=1.2cm}, 
		height=4cm, width=4cm,
	ylabel style={yshift=-0.1cm},
	every tick label/.append style={font=\small}
	]

\nextgroupplot[
ylabel=loss on Diginetica,
xlabel=epoch,
xmin=0,
xmax=11,
ymin=1.0,
ymax=2.5,
ytick={1.2,1.4,1.6,1.8,2.0,2.2,2.2,2.4},
ytick pos=left
]
\addplot coordinates {
(1,2.11778768)
(2,1.60713453)
(3,1.49758504)
(4,1.31929717)
(5,1.28020798)
(6,1.26028309)
(7,1.22622480)
(8,1.22226427)
(9,1.21954700)
(10,1.21542739)
};

\nextgroupplot[
ylabel=loss on Gowalla,
xlabel=epoch,
xmin=0,
xmax=11,
ymin=0.30,
ymax=1.3,
ytick={0.4, 0.6, 0.8, 1.0, 1.2},
ytick pos=left
]
\addplot coordinates {
(1,0.98434843)
(2,0.91657076)
(3,0.84457967)
(4,0.63606421)
(5,0.56618554)
(6,0.52026554)
(7,0.46130047)
(8,0.45455902)
(9,0.44974996)
(10,0.44330393)
};

\nextgroupplot[
ylabel=loss on Amazon,
xlabel=epoch,
xmin=0,
xmax=11,
ymin=0.2,
ymax=0.45,
ytick={0.2, 0.25, 0.30, 0.35, 0.40, 0.45},
ytick pos=left
]
\addplot coordinates {
(1,0.39160298)
(2,0.29188150)
(3,0.27538939)
(4,0.24992270)
(5,0.24476711)
(6,0.24248773)
(7,0.23831853)
(8,0.23793195)
(9,0.23766575)
(10,0.23717409)
};

\end{groupplot}

\end{tikzpicture}\vspace{-2mm}
   \caption{The convergence of our model.}\vspace{-2mm}
    \
\label{fig:conver}
\end{figure*}

Besides, to further verify the convergence of our model, we present how the loss of our model varies over epochs during the training process on the three datasets. Figure \ref{fig:conver} shows the experiment results, and we can observe that the model consistently converges.

\subsection{Case Study on Explanation Task (RQ4)}
Since CGSR has distinguished causality from correlation relationship between items, it is expected to facilitate the explanation task in SR in a fine-grained fashion.
In this case, we design an explainable framework on CGSR to clarify why a specific item $v_j\in \mathcal{V}$ is recommended given session $S$ on both session and item levels by generating a set of explanation scores. Specifically, in the \emph{explainable framework}, \textbf{on the session level}, three scores leading to the final recommendation score of $v_j$ (i.e., $Score_j^{ca},Score_j^r,Score_j^p$) are output by \emph{Recommendation Score Generator} in CGSR. \textbf{On the item level}, we calculate the score (importance) of each item $v^S_i$ in $S$ to $v_j$ under each relationship type (i.e., $Score_{ij}^{ca}$ for causality relationship and $Score_{ij}^r$ for correlation relationship):
\begin{equation}
\small
\begin{aligned}
&Score_{ij}^{ca}=(\mathbf{W_{c,6}} [\mathbf{x}_{i,S}^c; \mathbf{x}_{i,S}^c])^T \mathbf{x}_j^e-\gamma_1 (\mathbf{W_{e,6}} [\mathbf{x}_{i,S}^e; \mathbf{x}_{i,S}^e])^T) \mathbf{x}_j^c \\ 
&Score_{ij}^r=(\mathbf{W_{r,6}} [\mathbf{x}_{i,S}^r; \mathbf{x}_{i,S}^r])^T \mathbf{x}_j^r
\end{aligned}\label{eq:caseStudy}
\end{equation} 

Consequently, with the framework, given a recommended item $v_j$ and session $S$, we can not only understand the impact of $S$ under each type of relationship on $v_j$ from the session level, but also recognize the importance of each item in $S$ as either cause item or correlation item to $v_j$ from the item level.

\begin{figure*}[htbp]
    \centering
    \includegraphics[width=11cm]{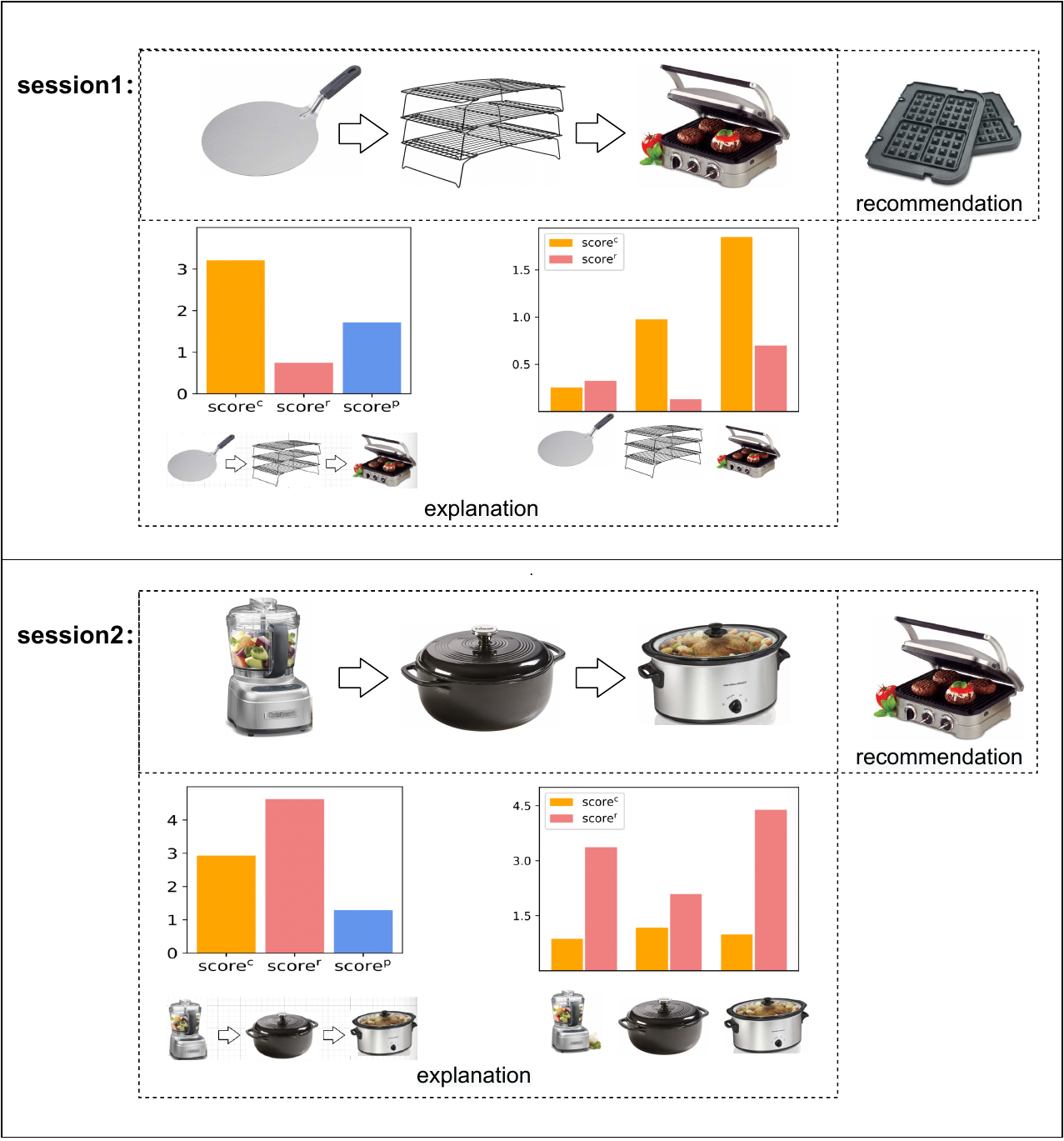}
    \caption{Two sessions \{Cake Lifter, Cooling Rack, Griddler, Griddler Waffle Plates\} and \{Mini-Prep Plus Food Processor, Oven with Dual Handles, Slow Cooker, Griddler\}, and the corresponding explanation scores on both session level and item level. }
    \label{fig:caseStudy}
\end{figure*}

In our experiment, to showcase the effectiveness of our explainable framework, we instantiate it on Amazon dataset. We choose Amazon dataset because item information on Amazon is publicly available while on other two datasets it is anonymously encoded.
Figure \ref{fig:caseStudy} depicts two cases on Amazon, which relate to two randomly chosen sessions: $S_1$ \{Cake Lifter, Cooling Rack, Griddler, Griddler Waffle Plates (recommended item)\} and $S_2$ \{Mini-Prep Plus Food Processor, Oven with Dual Handles, Slow Cooker, Griddler (recommended item)\}. The histogram on the left side presents the explainable scores on session level, whilst the histogram on the right side shows those on item level.

As shown in Figure \ref{fig:caseStudy}, for $S_1$, causality score $Score^{ca}$ is higher than correlation score $Score^r$, implying that on session level, Cake Lifter, Griddler and Cooling Rack in $S_1$ are more like cause items leading to buy Griddler Waffle Plates. On item level, under causality relationship, the score of Griddler is higher than the other two items, which means that Griddler is the main reason motivating to also purchase the Griddler Waffle Plates. Towards $S_2$, correlation score is higher than causality score. For each item in $S_2$, the corresponding correlation score with Griddler is also higher than the causality score. This is consistent with reality that Griddler has a correlation relationship rather than a causality relationship with Mini-Prep Plus Food Processor, Oven with Dual Handles, and Slow Cooker, respectively. 
From the two case studies, we can see that our explainable framework can provide valuable and reasonable explanations towards recommendations in SR.

\section{Conclusions}
\label{sec:conclusions}

The directed causality relationship between items, which has been ignored by previous studies, is quite important for effective session-based recommendation (SR).
In this paper, we proposed a novel method denoted as CGSR to explicitly consider causality and correlation relationship in SR. 
Specifically, on the basis of sessions, we constructed a cause graph, an effect graph, and a correlation graph considering both first-order and three types of second-order relationship, which are fed into a GNN and attention mechanism-based model to obtain four types of session representations for recommendation. 
By doing this, we can capture both causality and correlation relationship between items, and maximize item transition from cause to effect whilst simultaneously minimize that from effect to cause.
Extensive experimental results on three real-world datasets firstly revealed the superiority of our model over the state-of-the-arts, validating the effectiveness of distinguishing causality and correlation relationship.  
Secondly, exhaustive ablation studies verified the effectiveness of every component in CGSR. Thirdly, we also investigated the sensitivity of hyper-parameters and demonstrated the convergence of our model.
We further designed an explainable framework on CGSR to improve the explainability of SR. Case studies on Amazon dataset showcased that our framework can facilitate the explanation task in session-based recommendation.

For future work, we will continue to explore the causality and correlation relationship among items and even mine other relationship types for further improving recommendation accuracy. Besides, we aim to reveal the underlying factors leading to these directed item relationships by using popular causal inference techniques. Moreover, we strive to design other studies to provide the corresponding explainability.

\section{Acknowledgments}
We greatly acknowledge the support of Shanghai Rising-Star Program (Grant No. 23QA1403100), the National Natural Science Foundation of China (Grant No. 72192832), and the Natural Science Foundation of Shanghai (Grant No. 21ZR1421900).

\bibliographystyle{ACM-Reference-Format}
\bibliography{references}



\end{document}